\documentclass[letterpaper,twocolumn,10pt]{article}
\usepackage{usenix2019_v3}

\usepackage{epsfig,endnotes}
\usepackage{algorithmic}
\usepackage[linesnumbered,ruled]{algorithm2e}
\usepackage{hyperref}

\usepackage{mathtools}
\usepackage{graphicx}
\usepackage[capitalize]{cleveref}
\usepackage{xspace}
\usepackage[font=footnotesize,labelfont=bf]{caption}
\usepackage{textcomp}
\usepackage{comment}
\usepackage{booktabs}
\usepackage{subcaption}
\usepackage{fancyhdr}
\usepackage{todonotes}
\usepackage[htt]{hyphenat}
\usepackage[scaled=0.8]{beramono} %
\usepackage{listings}
\usepackage{xcolor}
\usepackage{authblk}

\definecolor{mygreen}{rgb}{0.1, 0.5, 0.1}

\usepackage{newunicodechar}
\newunicodechar{λ}{\ensuremath{\lambda}}
\newunicodechar{≈}{\ensuremath{\approx}}

\usepackage[compact]{titlesec}

\usepackage{enumitem}
\setitemize{noitemsep,topsep=0pt,parsep=0pt,partopsep=0pt}
\setenumerate{noitemsep,topsep=0pt,parsep=0pt,partopsep=0pt}
\setdescription{itemsep=0pt,topsep=0pt,parsep=0pt,partopsep=0pt,itemindent=0em,leftmargin=0.5em}

\newcommand{\us}{\textmu{}s\xspace}

\newenvironment{denseitemize}{
\begin{itemize}[topsep=2pt, partopsep=0pt, leftmargin=1.5em]
  \setlength{\itemsep}{2pt}
  \setlength{\parskip}{0pt}
  \setlength{\parsep}{0pt}
}{\end{itemize}}

\newenvironment{denseenum}{
\begin{enumerate}[topsep=2pt, partopsep=0pt, leftmargin=2em]
  \setlength{\itemsep}{2pt}
  \setlength{\parskip}{0pt}
  \setlength{\parsep}{0pt}
}{\end{enumerate}}

\def\ie{{i.e.\ }}
\def\eg{{e.g.,\ }}
\def\etal{{et al.\xspace}}

\usepackage{authblk}
\makeatletter
\renewcommand\AB@affilsepx{, \protect\Affilfont}
\makeatother

\usepackage{tikz}

\usepackage{framed}

\def\name{BPF-oF\xspace}
\def\syscall{\texttt{read\_bpfof}\xspace}

\newcommand{\tailpct}[0]{99\textsuperscript{th}-percentile}

\iftrue
\newcommand{\asaf}[1]{\textcolor{red}{\{Asaf: #1\}}}
\newcommand{\yannis}[1]{\textcolor{red}{\{Yannis: #1\}}}
\newcommand{\amy}[1]{\textcolor{red}{\{Amy: #1\}}}
\newcommand{\yuhong}[1]{\textcolor{olive}{\{Yuhong: #1\}}}
\newcommand{\junfeng}[1]{\textcolor{red}{\{Junfeng: #1\}}}
\newcommand{\ryan}[1]{\textcolor{red}{\{Ryan: #1\}}}
\newcommand{\hubertus}[1]{\textcolor{red}{\{Hubertus: #1\}}}
\newcommand{\jonas}[1]{\textcolor{red}{\{Jonas: #1\}}}
\newcommand{\sheng}[1]{\textcolor{cyan}{\{Sheng: #1\}}}
\newcommand{\kostis}[1]{\textcolor{red}{\{Kostis: #1\}}}

\newcommand{\tanvir}[1]{\textcolor{brown}{\{TAK: #1\}}}

\else
\newcommand{\asaf}[1]{}
\newcommand{\yannis}[1]{}
\newcommand{\amy}[1]{}
\newcommand{\yuhong}[1]{}
\newcommand{\junfeng}[1]{}
\newcommand{\ryan}[1]{}
\newcommand{\hubertus}[1]{}
\newcommand{\jonas}[1]{}
\newcommand{\kostis}[1]{}
\newcommand{\sheng}[1]{}
\newcommand{\tanvir}[1]{}
\fi

\begin{document}
\date{\vspace{-1.3cm}}

\title{\name: Storage Function Pushdown Over the Network}

\author[*,1]{Ioannis Zarkadas}
\author[*,1]{Tal Zussman}
\author[1]{Jeremy Carin}
\author[1]{Sheng Jiang}
\author[1]{Yuhong Zhong}
\author[3]{Jonas Pfefferle}
\author[3]{Hubertus Franke}
\author[1]{Junfeng Yang}
\author[1]{Kostis Kaffes}
\author[2]{Ryan Stutsman}
\author[1]{Asaf Cidon}
\affil[1]{Columbia University}
\affil[2]{University of Utah}
\affil[3]{IBM}
\affil[*]{denotes equal contribution}

\maketitle

\begin{abstract}

Storage disaggregation, wherein storage is accessed over the network,
is popular because it allows applications to independently scale storage capacity and bandwidth based on dynamic application demand.
However, %
the added network processing introduced by disaggregation can consume significant CPU resources. %
In many storage systems, logical storage operations (\eg lookups, aggregations) involve a series of simple but dependent I/O access patterns. %
Therefore, one way to reduce the network processing overhead is to execute dependent series of I/O accesses at the remote storage server, reducing the back-and-forth communication between the storage layer and the application. We refer to this approach as \emph{remote-storage pushdown}.
We present \name, a new remote-storage pushdown protocol built on top of NVMe-oF, which enables applications to safely push custom eBPF storage functions to a remote storage server. %

The main challenge in integrating \name with storage systems is preserving the benefits of their client-based in-memory caches. %
We address this challenge by designing novel caching techniques for storage pushdown, including splitting queries into separate in-memory and remote-storage phases and periodically refreshing the client cache with sampled accesses from the remote storage device. %
We demonstrate the utility of \name by integrating it with three storage systems, including RocksDB, a popular persistent key-value store that has no existing storage pushdown capability.
We show \name provides significant speedups in all three systems when accessed over the network, for example improving RocksDB's throughput by up to 2.8$\times$ and tail latency by up to 2.6$\times$. %

\end{abstract}

\section{Introduction}

Storage disaggregation enables accessing block storage devices over the network and scaling independently of the application~\cite{reflex,flash-disagg,nvme-of-disagg,gimbal,rack-level-storage,hailstorm,decibel}.
As datacenter networks become faster ($\sim$10~\textmu{}s latency), accessing storage devices quickly over the network becomes feasible. Similarly, datacenter networks offer ample bandwidth to saturate the I/O bandwidth of storage devices~\cite{reflex,flash-disagg,gimbal}.
Consequently, a variety of data-intensive datacenter applications (\eg transactional databases~\cite{aurora,cornus}, data analytics~\cite{rockset}, key-value stores~\cite{rocksdb-evolution}, and data warehouses~\cite{snowflake}) use storage disaggregation.

NVMe-oF (NVMe over Fabrics) is the standard networked storage protocol for storage disaggregation. NVMe-oF allows an application to transparently access an NVMe block device on a remote server. NVMe-oF supports TCP and RDMA, is fully supported by Linux, and can be used without specialized hardware.
The major downside of NVMe-oF (compared to accessing a device locally via NVMe) is the added network latency and the CPU cost of processing the network packets. %
The CPU cost is especially substantial in the case of TCP, incurring a nearly 50\% throughput decrease compared to local performance in our experiments (\S\ref{sec:motivation-experiment}).
We show that even in the case of RDMA, where this cost can be partially mitigated by offloading network processing to NICs, network processing still incurs a significant toll.
With the goal of reducing this cost, we observe that in modern storage systems, many queries, such as key-value lookups and aggregations, involve a series of dependent I/O accesses (\eg a B-tree index lookup), which translate into multiple round-trip accesses in the disaggregated setup.
Therefore, one approach to amortize the network processing cost of storage disaggregation is to ``pushdown'' dependent I/O accesses to the remote storage server, thereby eliminating the need to process multiple round-trip network requests for each logical query. This idea itself is not new, and there are many examples of academic and commercial systems that push down functions over the network to operate closer to the data~\cite{kourtis2020safe,splinter,kayak,adaptive-placement,lefevre2020skyhookdm,amazon-select}. However, these systems all implement this capability as an \emph{application-specific feature}.
Restructuring a storage system or database to allow it to run its operations on a remote server may entail many intrusive modifications, and it requires developers to contend with how to implement such a capability while maintaining security and concurrency guarantees.

We take a cue from recent work~\cite{xrp,kourtis2020safe,lambda-io,wu2021bpf} that has explored the use of eBPF (extended Berkeley Packet Filter) as a general-purpose framework for safely executing user-defined and kernel-verified storage functions closer to the storage device within the Linux NVMe driver~\cite{xrp,wu2021bpf} or a programmable storage device~\cite{lambda-io}, thereby reducing the CPU overhead of traversing the kernel storage stack.
However, these prior works assume pushdown occurs locally (\ie on the same server) or on simple storage systems.

Two significant challenges arise when trying to adopt a similar approach in a disaggregated storage setting with a modern storage system.
First, due to the design of block-level protocols like NVMe-oF, the programmable function that executes on the remote server operates below the file system layer (\ie at the NVMe driver). However, user-defined storage functions, such as index traversals, operate on files. Prior work on local eBPF storage pushdown~\cite{xrp} has solved this by synchronously synchronizing the file-to-block mappings between the file system and NVMe driver. However, such an approach would be very slow for a networked setting, as all storage functions would block until the metadata synchronizes between the client and server.
Second, in modern storage systems, data is stored in large data structures (\eg LSM-trees or B-trees), and parts of these data structures may be partially (or fully) cached in memory and are frequently updated. However, in a networked setting, these in-memory structures would not be accessible at the remote server, which runs the I/O functions. Thus, if we use remote-storage pushdown, we may lose the benefit of these in-memory caches, hurting performance.

We solve these challenges by designing \name, a new \emph{general-purpose} remote-storage pushdown protocol for TCP and RDMA that is built on top of NVMe-oF, allowing a client to submit custom eBPF storage functions and safely trigger them at a remote storage server. %
\name's design relies on a key observation: modern storage systems often know in advance which files they might access to satisfy a storage operation. Thus, we tackle the first challenge by requiring the application to specify upfront which files it will access throughout its storage operation. \name uses this information to perform the file-descriptor-to-inode translation at the client.
To keep the file-to-block mappings up to date, \name updates them asynchronously, thereby avoiding blocking concurrent storage functions. %
To avoid conflicts when storage functions execute concurrently with metadata synchronization, \name applies a version to each inode's file-to-block mapping, and it checks for modified versions before issuing and after completing a request.
We implement \name as a new NVMe command that is triggered by a system call from userspace, which also registers the required eBPF functions for the storage pushdown.

We tackle the second challenge using a new technique called
\emph{query splitting}, which divides queries into an in-memory portion and an on-disk portion. %
Accessing in-memory data structures is often orders of magnitude cheaper than a single storage I/O. Therefore, the in-memory portion of the query speculatively checks all the potentially-required in-memory (cached) data structures at the client (some of which may eventually not be needed), and then, based on the results, executes the remaining on-disk portion of the query at the remote server. %
While this approach may result in ``unnecessary'' accesses to in-memory data structures (\eg looking up bloom filters for files that will never be used), it makes the on-disk portion of the query much more efficient by grouping all I/O accesses together.
Another caching-related problem is that the client cache may become stale, as it is not updated with the data that is accessed by the remote function. One potential solution to this problem is to simply return the intermediate data accessed by the remote storage query, but this would increase the network consumption between the client and remote server, negating much of the benefit of \name. Instead, %
we solve this problem using a lightweight technique of \emph{cache sampling}, where a small percentage (by default 1\%) of storage functions do not get offloaded, and fully update the client cache, keeping it fresh. %

Our experiments yield another surprising result: for some workloads, maintaining a client data cache with \name actually \emph{hurts} performance, because \name is very CPU efficient and the cost of maintaining the client cache is higher than its benefit. For example, when we run \name with a uniform workload on fast storage devices, it is in fact better to use \name with no cache (beyond indices) at the client!

We integrate \name with three storage systems: RocksDB~\cite{rocksdb} a popular and performance-optimized storage system used by Meta and many other companies~\cite{rocksdb-users}, BPF-KV~\cite{xrp}, a toy eBPF-friendly key-value store, and WiredTiger~\cite{wiredtiger}, a simple storage engine used by MongoDB. Our integration required a relatively modest change to RocksDB and WiredTiger: about 1,900 and 500 lines of code, respectively. \name provides significant performance boosts to all three systems: under the vast majority of workloads \name accelerates RocksDB by up to 2.8$\times$ and reduces \tailpct ~latency by up to 2.6$\times$; it accelerates BPF-KV by up to 8$\times$; and it improves WiredTiger's throughput by up to 30\%.
These gains are most apparent in read-heavy workloads and workloads where the working set does not fit in cache, but they translate to other workloads as well. However, as expected, the utility of storage pushdown is lower in write-heavy workloads or when almost all requests are serviced from in-memory data structures.
We will make \name and our storage system integrations open source upon publication.
Our major contributions are:
\begin{denseitemize}
\item \name, an efficient mechanism to safely enable storage pushdown for storage applications running on disaggregated storage. \name is the first implementation of storage pushdown in a general networked storage protocol, and its key ideas are an efficient and asynchronous file metadata synchronization mechanism and a versioning scheme for inode mapping to protect concurrent access to files. \name greatly reduces the CPU and network overhead of NVMe-oF while requiring no custom hardware.
\item Our novel query splitting and cache sampling techniques, which allow \name to support local in-memory caching, address a significant challenge in storage pushdown. We integrate \name with modern storage systems, including the highly-optimized RocksDB.
\end{denseitemize}

\section{Background and Motivation}

This section first provides a background on NVMe-oF. %
It then discusses the CPU overhead of NVMe-oF, providing the motivation for remote storage pushdown. Finally, we review eBPF as a framework
for storage pushdown.

\subsection{NVMe-oF Primer}
\label{sec:nvmeof-primer}

\begin{figure}[t]
    \centering
    \includegraphics[width=\columnwidth]{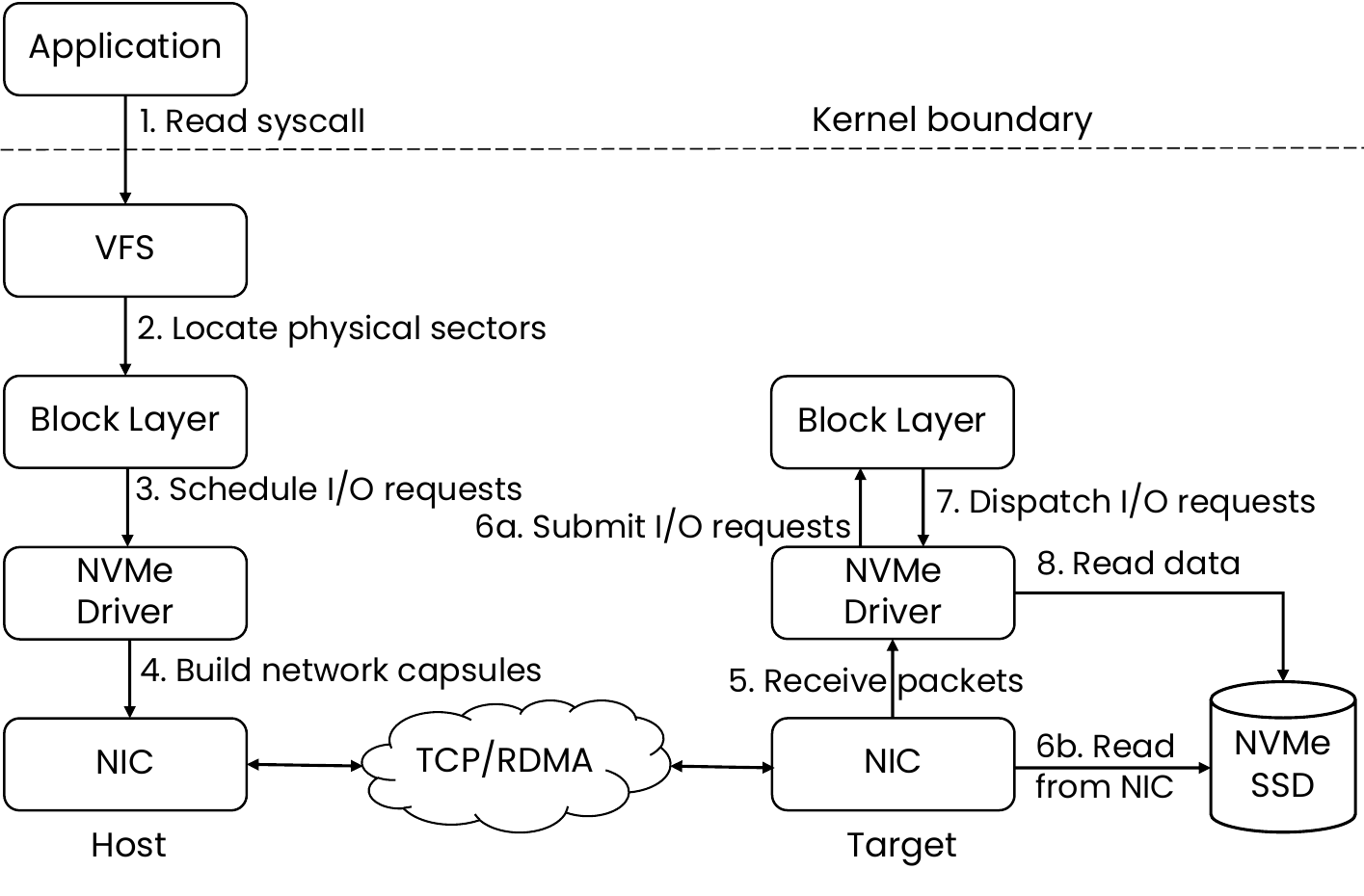}
    \vspace{-2em}
    \caption{NVMe-oF overview.}
    \label{fig:nvmeof-overview}
    \vspace{-1.5em}
\end{figure}

NVMe-oF has emerged as the main networked storage protocol for disaggregated storage, replacing iSCSI due to its better performance~\cite{nvme-of-disagg,iscsi-nvmeof}. NVMe-oF allows an application to directly access a block storage device connected to a remote server using NVMe. %
Figure~\ref{fig:nvmeof-overview} shows the flow of an NVMe-oF read request. The \emph{host} (\ie the client, left) initiates the request, and the \emph{target} (\ie the server, right) contains the SSD and performs the actual disk I/O. To initiate an NVMe-oF request, an application at the host issues a storage system call, such as \textsc{read} (step 1), which then traverses its local OS storage stack (steps 2-3). The request is treated as a regular NVMe request until it reaches the local NVMe driver.
The \emph{host} and \emph{target} drivers maintain I/O queues for exchanging the NVMe-oF \emph{capsule}, which is the unit of communication between the host and the target (\ie a capsule is to NVME-oF what a frame is to Ethernet). The NVMe driver handles the request by constructing an NVMe-oF command within a capsule and submitting it to an NVMe I/O queue. The capsule is then forwarded to the relevant network stack (step 4) depending on the fabric type (TCP, RDMA, Fiber Channel), and it is transmitted to the target.
At the target (step 5), after the driver extracts the NVMe-oF command, it generates the block layer request and submits it to the block layer for I/O scheduling (step 6a). The target's NVMe driver, at last, receives the I/O request from the block layer (step 7) and reads the user's data from the local NVMe SSD (step 8), which is then sent back to the host via the fabric-specific mechanism (not shown in the figure). %
Major NIC model lines (\eg NVIDIA ConnectX, Broadcom Stingray, Intel IPU) support offloading the NVMe-oF target datapath to the NIC when the underlying protocol is RDMA, and some NICs even support TCP offload~\cite{chelsio}. This allows the NIC to directly read from or write to the NVMe device, bypassing the target's CPU (step 6b). %

\begin{figure}[!t]
    \centering
    \begin{subfigure}[t]{0.23\textwidth}
        \includegraphics[width=\columnwidth]{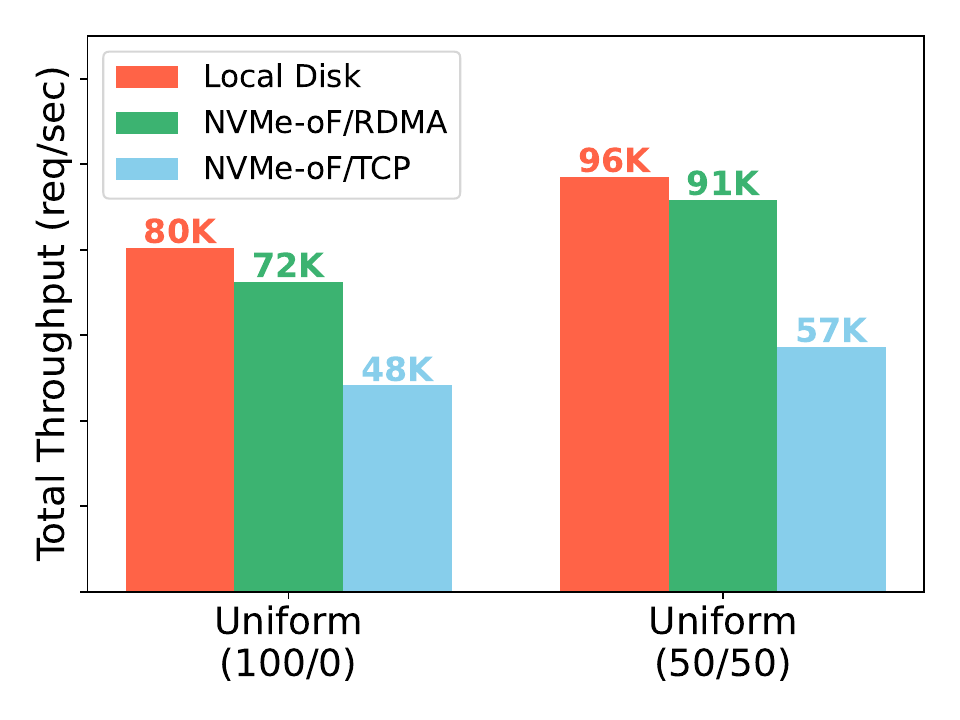}
        \vspace{-2em}
        \caption{Throughput.}
    \end{subfigure}
    \hfill
    \begin{subfigure}[t]{0.24\textwidth}
        \includegraphics[width=\columnwidth]{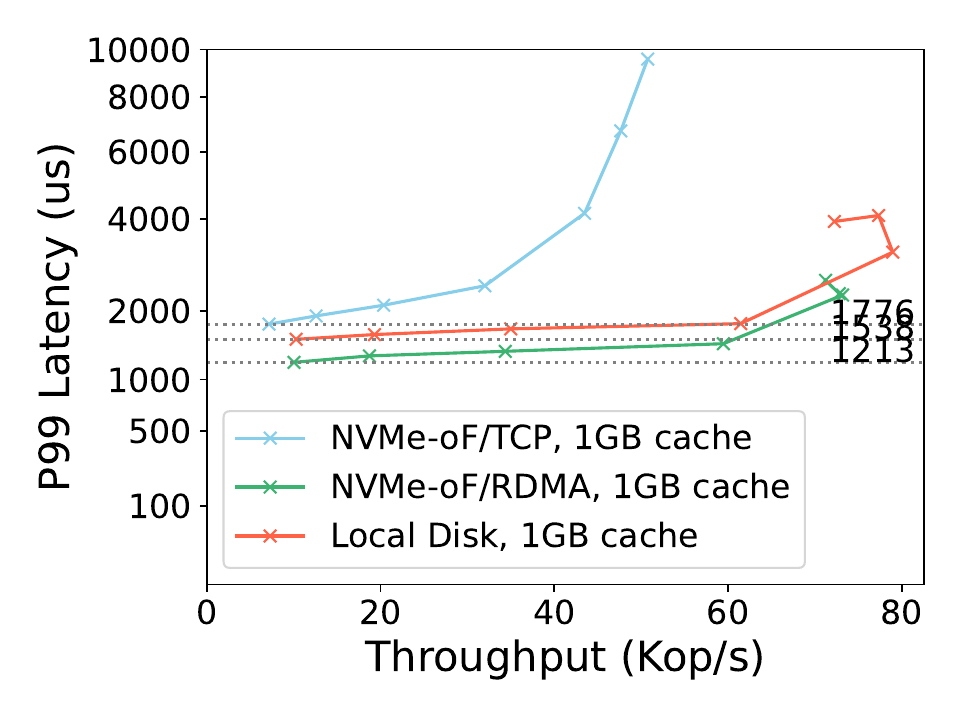}
        \vspace{-2em}
        \caption{Throughput vs. latency.}
    \end{subfigure}
    \vspace{-1.25em}
    \caption{RocksDB run on local disk compared to NVMe/RDMA (with offloading) and NVMe/TCP on a NAND SSD.}
    \label{fig:rocksdb-local-vs-nvmeof-tcp}
    \vspace{-1.5em}
\end{figure}

\subsection{Improving NVMe-oF CPU Efficiency}
\label{sec:motivation-experiment}

A significant drawback of NVMe-oF is the CPU overhead of processing the packets, both at the host and the target~\cite{nvme-of-disagg}. This is especially acute in the case of NVMe/TCP, due to the relatively high CPU cost of processing TCP packets~\cite{reflex,i10,demikernel}.
High CPU consumption can adversely affect the performance of data-intensive applications, especially when they are CPU-bound.

To demonstrate the network processing cost of NVMe-oF, we compare running two workloads on a 100~GB RocksDB instance with a NAND SSD locally to NVMe/RDMA (with NIC offloading) and NVMe/TCP (Figure~\ref{fig:rocksdb-local-vs-nvmeof-tcp}).
Both workloads request objects by keys drawn uniformly at random: the first is read-only, and the second is 50\% reads, 50\% updates. Each RocksDB instance uses 1~GB of cache and 3 cores locally at the host. The remote target is also allocated 3 cores, which are only partially used by the NVMe/TCP setup. Our experimental setup on CloudLab is fully described in \S\ref{sec:experimental-setup}.

By running on a local NAND SSD, RocksDB achieves a 1.5--1.6$\times$ throughput improvement compared to running on the same NAND SSD accessed via NVMe/TCP. While using RDMA is more CPU-efficient than TCP, as expected, NVMe/RDMA is still 12--15\% slower than the local disk. The primary cause of NVMe-oF's slowdown is the CPU overhead of processing the packets at the host, as RocksDB is often bottlenecked by CPU. In addition, NVMe-oF/TCP consumes additional CPU cycles at the target. Similarly, running RocksDB with NVMe/TCP incurs a significant latency overhead.
While NVMe-oF provides significant operational advantages, it incurs a non-negligible performance overhead due to the extra network processing cost and communication time. %

\vspace{-0.5em}\paragraph{Motivation for storage pushdown.}
A promising way to reduce the CPU and network consumption of NVMe-oF is to ``push down'' computation to the target storage server. This reduces potential back-and-forth communication between the target and host.
Storage function pushdown has been explored extensively in the database community as ``predicate pushdown''~\cite{hellerstein1993predicate,levy1994query,flexpushdown}, in the systems community as near-storage compute~\cite{lambda-io,splinter,kayak,adaptive-placement,lefevre2020skyhookdm} and, more recently, as user-defined functions in the kernel~\cite{xrp, kourtis2020safe,wu2021bpf,extfuse}. For more details, see \S\ref{sec:related}. However, to the best of our knowledge, no prior work has combined storage function pushdown with a general-purpose networked storage protocol such as NVMe-oF.

Key-value stores such as RocksDB are amenable to such an approach for two reasons.
First, many RocksDB read operations are composed of a sequence of dependent I/O operations.
For example, in the previous experiment, RocksDB issues on average 3.8 I/Os and 4 I/Os per read request, for the read-only and 50-50 read-write workload, respectively. Most read requests must access multiple files, since files are spread across multiple levels of RocksDB's on-disk tree data structure. %
Second, RocksDB only writes data to disk in large immutable files. This significantly simplifies synchronization between potentially-conflicting simultaneous reads and writes. We will expand on this point later in the paper.
Since many SSD-optimized storage systems exhibit these two properties~\cite{wiredtiger,leveldb,pebblesdb,wisckey,cassandra,rockset,taft2020cockroachdb,sqlite}, pushing computation closer to the device can alleviate the overheads of accessing storage devices over the network for a wide variety of deployments.

\subsection{Local eBPF-based Storage Pushdown}

The idea of storage pushdown can be applied locally within a single host. XRP~\cite{xrp} is an eBPF framework that adds a hook to the Linux kernel's NVMe driver, which allows installation of application-defined functions that can intercept NVMe I/O completions directly in the kernel's NVMe interrupt handler. This allows these application-defined functions to immediately process the data returned after each block I/O access. These functions can process, aggregate, or filter the results before they are returned to the application. These functions can also efficiently trigger additional block I/Os by recycling the NVMe I/O descriptor to read additional blocks based on the data they observed in earlier I/O completions. For example, application-defined functions can traverse a B-tree stored in a file and return a requested key-value pair with a single system call, despite the multiple, data-dependent I/Os needed for the traversal.
Using these application-defined functions, XRP showed that data-dependent operations like index traversal in two simple key-value stores (BPF-KV and WiredTiger) led to significant speedups even in a single host.
However, these speedups were demonstrated with simple applications: locally-run key-value stores where all I/O operations involve one file with one cache and static fixed-sized data structures on disk. %

\vspace{-0.5em}\paragraph{Why eBPF?} \name also uses eBPF to enable applications to run custom storage pushdown functions.
By running the application-defined storage functions as eBPF functions, one can ensure that they are isolated from one another and from sensitive kernel state.
In order to safely allow user functions to execute in the kernel, eBPF functions are statically verified by the kernel to ensure several safety properties. For example, functions cannot contain an unbounded number of instructions, and they cannot access memory addresses outside the sandbox without the help of special, trusted helper functions that the kernel makes available to eBPF functions. eBPF functions are efficient partly because they can eliminate the need to switch to userspace and back to run user-defined code, and because eBPF functions are JIT-compiled to native machine code by the kernel when they are registered with a specific kernel hook site (such as the aforementioned NVMe hook). %

\section{Challenges}
\label{sec:challenges}

In this section, we describe the main research challenges in allowing storage functions to be triggered over the NVMe-oF protocol on a modern storage system. %

\vspace{-0.5em}\paragraph{Challenge 1: Safely access files in the target.}
As described in \S\ref{sec:nvmeof-primer}, in the NVMe-oF protocol, %
once the request reaches the target, it only knows how to access blocks, and does not have any file system information, which resides at the host. Therefore, an eBPF function that submits storage I/O for a file on the target must determine the location of the next block without having access to the file system metadata mapping files to blocks. In addition, our system must support resubmission across different file descriptors (\eg when traversing an LSM-tree) to support modern storage applications. Thus, the target must be able to translate a file descriptor to the corresponding inode and then find the block offset corresponding to the requested file offset.
It must do so in a safe manner, without the risk of reading and returning invalid data.

A simpler version of this problem also existed in XRP's setting, where functions are resubmitted at the local NVMe driver, which sits below the file system. XRP solved this problem by synchronously maintaining a ``metadata digest'', which is an RCU-protected copy of the file system's files-to-blocks mapping~\cite{xrp}. The problem becomes more complex in the networked setting, because the ``metadata digest'' has to be maintained over the network, resubmissions can target different files, and \name must guarantee that resubmissions read and return valid data in the face of concurrent requests.

\vspace{-0.5em}\paragraph{Challenge 2: Integrate in-memory data structures.}
Integrating a production storage system with \name presents a number of interesting challenges, primarily owing to the interactions between the system's in-memory data structures, which sit locally at the host, with the remote function execution, which now occurs remotely at the target. First, when executing a storage operation, storage systems typically do not produce a ``pure'' chain of I/O accesses. Instead, some of the intermediate data structures in the operation path may be cached in memory. For example, in LSM-tree storage systems~\cite{leveldb,rocksdb,o1996log,splinterdb,pebblesdb} each file has its own index, and these indices are often cached in memory, while the actual data mostly resides on disk. Therefore, a naive pushdown operation that traverses multiple files will either ignore these cached indices, issuing many more storage I/Os than necessary, or return back to the host for each file, negating the benefit of storage pushdown.
In order to achieve optimal speedup with \name, we need to ideally chain the I/O storage accesses together in one network traversal.

Second, assuming we offload such a chain of I/O accesses, there is still the question of how to keep the host's in-memory cache up-to-date. While the final result of the query will always be returned, the ``intermediate'' I/O accesses (\eg accessing a file in the middle of the LSM-tree lookup) would not. A strawman approach would to be return the results of \emph{all} the I/O accesses back to the host and then appropriately update the application's in-memory data structures. However, this would negate much of the benefit of storage pushdown, as these intermediate results are not actually needed by the application (only the final query result is needed), and would lead to approximately the same network bandwidth consumption as running the requests with no storage pushdown.

\section{\name Design}

In this section we present \name's design principles (\S\ref{sec:principles}), \name's general architecture (\S\ref{sec:arch}), and discuss how it addresses the first challenge from \S\ref{sec:challenges}. Finally, we discuss the limitations of our current design in \S\ref{sec:limitations}. %

\subsection{Principles}
\label{sec:principles}

We design \name with the following principles in mind.

\begin{denseenum}
\item \textbf{Wide applicability.} Require minimal modifications to the NVMe-oF protocol, Linux kernel, and application that uses \name. In addition, the system should be immediately deployable on public clouds, and it should not require any specialized hardware support.
\item \textbf{CPU efficiency.} Increase throughput per core and reduce latency. %
\item \textbf{Concurrent access.} \name must not return invalid data in the face of concurrent requests that may modify file system metadata.
\item \textbf{Flexible resubmissions.} Support variable-sized resubmissions across different files, to enable data structure traversal in modern storage systems. %
\item \textbf{Target SSD-optimized applications.} We target SSD-optimized storage applications~\cite{wiredtiger,leveldb,pebblesdb,wisckey,cassandra,rockset,taft2020cockroachdb}, wherein incoming updates are written to in-memory buffers and are periodically flushed to disk as large immutable files. \name primarily focuses on supporting read-only eBPF functions, since writes are background operations. It is also beneficial that such applications do not issue in-place updates to disk, otherwise updates would interfere with pushdown operations.
In addition, \name initially only targets applications that implement their own userspace cache (\ie no support for the OS page cache), which is the case for many SSD-optimized systems.
\end{denseenum}

\subsection{Architecture}
\label{sec:arch}

\begin{figure}[t]
    \centering
    \includegraphics[width=\columnwidth]{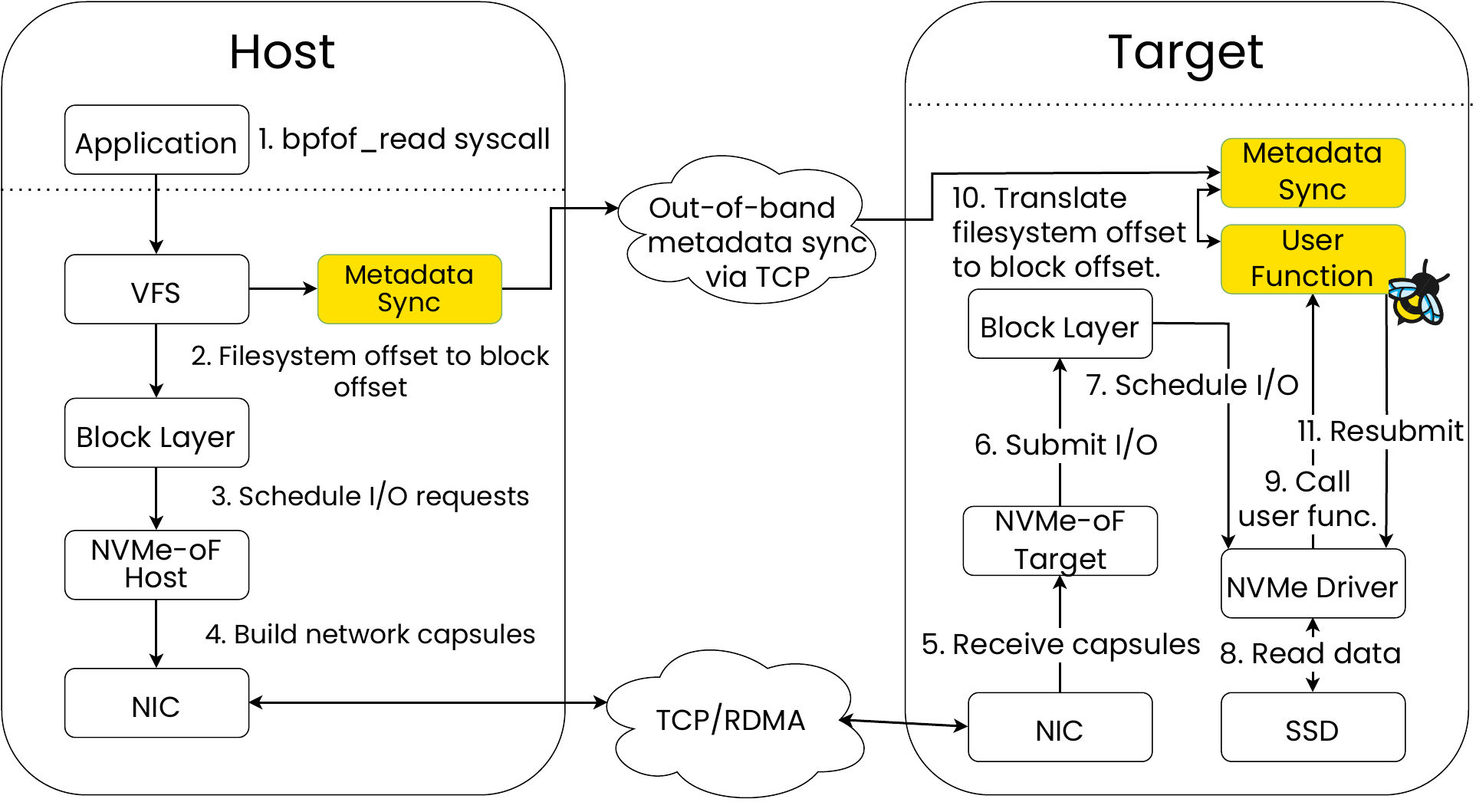}
    \caption{\name architecture.}
    \vspace{-1.5em}
    \label{fig:bpfof-architecture}
\end{figure}

\name supports storage pushdown with eBPF (Figure~\ref{fig:bpfof-architecture}) over NVMe-oF by adding three mechanisms.
First, it adds a mechanism to asynchronously update the target's file-to-block mapping. Second, it adds a versioning scheme that allows eBPF functions to safely access storage blocks at the target concurrently with file metadata updates. These two mechanisms jointly enable the third mechanism: safely running user-defined functions at the target server's NVMe driver.
Applications interact with \name using two interfaces: a new NVMe command that sends a ``pushdown'' NVMe-oF request and a userspace API that allows applications to install remote eBPF functions at the target.
Before diving into these mechanisms and interfaces, we walk through an example of a \name request that traverses a B-Tree index by pushing down the traversal to the target. We match the steps of the example to the steps shown in Figure~\ref{fig:bpfof-architecture}.

\vspace{-0.5em}\paragraph{End-to-end example of a \name request.}
At initialization, the application installs the remote eBPF index traversal function using \name's API. In our current implementation, the application on the host machine transfers the eBPF program over \textsc{scp} to the target machine, where it is compiled, verified and loaded. %
For each query, the application first stores the query key in a buffer, which is called the \textit{scratch buffer}~\cite{xrp}. The scratch buffer is allocated by the application and stores the query key and the query result from the remote eBPF function. In addition, the application specifies a list of file descriptors that it might access remotely. The first file descriptor in this list will be read first. The application also specifies the number of bytes and offset to read from the first file descriptor, and the eBPF function ID to call on the target.

Then, the application calls our \syscall system call (step 1). The system call obtains a reference for each passed file descriptor, ensuring that the files' data blocks are not overwritten if the file is deleted while there are outstanding requests. In addition, it creates a file-descriptor-to-inode mapping, which lets us avoid synchronizing the application's entire file-descriptor table by using inode versions. As described in~\S\ref{sec:versioning}, an inode's version changes whenever the file-to-block mapping (\eg extent tree) for that inode is updated. Then, the application's file system request is translated to a block request (step 2) and reaches the NVMe-oF host driver (steps 3, 4). There, the host constructs a new \syscall NVMe command. The command transfers the scratch buffer and its size to the target, along with the file-descriptor-to-inode mapping, the offset to read from, the number of bytes to read, and the eBPF function ID to call. Before submitting the request, the host driver checks that the inode versions have been synchronized, as described in Algorithm~\ref{algo:host-pre-check}.

Next, the NVMe-oF target driver receives the request (step 5), and prepares a block request and submits it (step 6). The block request is translated to an NVMe read request and submitted to the target's NVMe device (steps 7, 8).
The target's NVMe driver receives the request result and calls the user's eBPF function (step 9), passing in the result and the scratch buffer. The eBPF function processes the request result and decides what happens next. In this example, the returned data represents an internal node of the B-tree; hence, the eBPF function parses the node to find the file offset of the child node that it should read to continue the point query. The eBPF function indicates to \name's \emph{resubmission mechanism} that it should read this additional block, so \name resubmits the read request to the target's NVMe driver with the new offset (steps 10, 11). The resubmitted read's file descriptor is translated with the previously sent file-descriptor-to-inode mapping, and the offset is translated using the inode's file-to-block mapping, which the host synchronized with the target.

When the eBPF function finds the correct key, it writes the value to the scratch buffer and indicates that the result should be sent to the host. The scratch buffer is then returned to the host, where \name will again verify that the inode versions remain in sync, as shown in Algorithm~\ref{algo:host-post-check}. Finally, \name returns the result to the application.

In this way, \name implements index traversal at the target without requiring back-and-forth communication with the host. This workflow contains a versioning scheme, which allows \name to function correctly even in the face of concurrent reads and writes, which can lead to stale remote metadata. We now provide more details on these mechanisms and interfaces. For simplicity, in the description below, we assume an integration with the ext4 file system. %

\subsubsection{Mechanisms}
\label{sec:versioning}
\paragraph{Versioning.}

\name guarantees safety by implementing versioning on the file system metadata.
To do so, we make a minor change to the file system in order to add a version to each file's extent tree and increment it when the tree changes. With that functionality in place, we design our versioning algorithm, described by Algorithms~\ref{alg:versioning-before-submission} and~\ref{alg:versioning-after-completion}.

The versioning algorithm runs only on the host, before request submission and after completion.
Before submission, the host driver gets the latest inode ID and version for the given requests' file descriptors. For all of them, it checks if that version of the inode's extent tree has been synchronized with the target. If not, it aborts the request. This ensures that the target always has the latest version of the file system metadata. This check is actually enough to guarantee that the file-to-block mapping will be correct on the target, except for one case: when the extent tree is remapped, \ie an existing file offset points to a new disk offset. This can happen during truncation and defragmentation. We handle this case by adding a final version check when the response is received from the target: if the version has changed, the host aborts the request. It also wipes the scratch buffer to ensure that the target will not return stale data, even if the extent tree is remapped during the request.

\begin{algorithm}[t]
    \caption{Host - Check Before Submission}\label{alg:versioning-before-submission}
    \label{algo:host-pre-check}
    \begin{algorithmic}[1]
    \FOR{$idx = 0$ \TO $\text{num\_fds} - 1$}
        \STATE $fd \gets \text{bpfof\_request.fds}[idx]$
        \STATE $curr\_version \gets \text{get\_fd\_version}(fd)$
        \STATE $sent\_version \gets \text{latest version sent to target}$
        \IF{$curr\_version \neq sent\_version$}
            \STATE \textbf{abort request}
        \ENDIF
        \STATE $\text{request.fd\_versions}[idx] \gets curr\_version$
    \ENDFOR
    \end{algorithmic}
\end{algorithm}

\begin{algorithm}[t]
    \caption{Host - Check After Completion}\label{alg:versioning-after-completion}
    \label{algo:host-post-check}
    \begin{algorithmic}[1]
    \FOR{$idx = 0$ \TO $\text{num\_fds} - 1$}
        \STATE $fd \gets \text{bpfof\_response.fds}[idx]$
        \STATE $curr\_version \gets \text{get\_fd\_version}(fd)$
        \STATE $used\_version \gets \text{get\_sent\_request\_fd\_version}(\text{bpfof\_response})$
        \IF{$curr\_version \neq used\_version$}
            \STATE abort request
        \ENDIF
    \ENDFOR
    \end{algorithmic}
\end{algorithm}

\vspace{-0.5em}\paragraph{Metadata synchronizer.}
The next challenge is synchronizing the file system metadata (\ie file-to-block mappings) from the host to the target. The metadata synchronizer consists of two kernel modules running on the host and the target. The host module watches for file system changes by installing \emph{fsnotify} kernel hooks. For each change, it schedules a metadata synchronization for the specified inode. A kernel workqueue processes these requests and checks if the file-to-block mapping has changed from the last version it synchronized, using the versioning scheme described above. Then, the synchronizer copies the inode's extent tree and transfers it to the target module, where it is used for resubmissions. Metadata synchronization is done asynchronously over a TCP connection and does not block the data path. For simplicity, we implemented the synchronizer as an out-of-band TCP client and server, rather than creating a new NVMe command for it.

\vspace{-0.5em}\paragraph{eBPF execution.}
We design the eBPF execution and resubmission mechanism at the target on top of XRP~\cite{xrp}. We add support for (a) resubmitting eBPF functions across multiple files instead of a single file and (b) specifying a different request size for each resubmission.
We provide more details on these changes in \S\ref{sec:rocksdb}.

\subsubsection{Interfaces}

To integrate with NVMe-oF, \name implements a new NVMe command following the specifications for NVMe and NVMe-oF~\cite{nvme-spec,nvmeof-spec}. %
The command transfers the scratch buffer, which contains the query key and result (on completion), from the host to the target and back. The command's arguments are the same as a read request. We transfer the additional arguments needed by \name, as described above, with NVME-oF's data transfer mechanism (scatter-gather lists). %

\begin{figure}[t]
    \centering
    \includegraphics[width=\columnwidth]{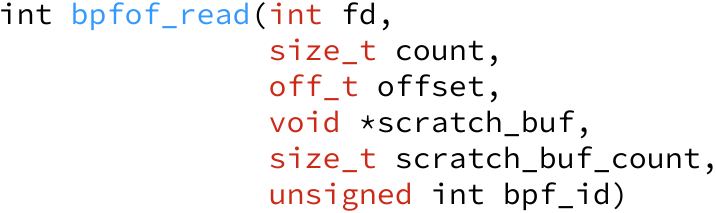}
    \vspace{-1.5em}
    \caption{\name read system call.}
    \vspace{-1em}
    \label{fig:bpfof-read-syscall}
\end{figure}

\name implements the new \syscall system call (Figure~\ref{fig:bpfof-read-syscall}) with the following arguments: a list of file descriptors it might read from, the number of bytes and offset to read from the first file descriptor, the scratch buffer and its size, and the eBPF function ID to call on the target. %

\subsection{Limitations}
\label{sec:limitations}

We view the current system as the first step in enabling pushdown in disaggregated storage settings. \name's current design has several limitations, which we discuss below, and plan to address in future work.

\vspace{-0.5em}\paragraph{Files known in advance.}
\name assumes that all files accessed in an I/O chain are known in advance. This assumption holds true in many storage systems (\eg LSM-tree storage systems), but may not hold in others. %

\vspace{-0.5em}\paragraph{Block device support.}
Currently, \name runs eBPF functions within the target's NVMe driver. From a performance standpoint, this is optimal, as it allows \name to almost entirely bypass the target kernel's storage stack, thereby reducing CPU consumption at the target. However, this design choice has several disadvantages: most notably, it doesn't support complex block device configurations (\eg RAID, LVM) but only single partition drives. However, \name could run the eBPF functions at a higher layer of the kernel (\ie in the target's block layer) in exchange for a small performance hit. %

\vspace{-0.5em}\paragraph{File system support.}
We currently only support the ext4 file system. %
We anticipate the integration for additional file systems will be very similar to ext4. %

\vspace{-0.5em}\paragraph{Limitations of eBPF.}
While eBPF has several attractive properties, including its usage in popular systems~\cite{cloudflare-ebpf,cilium}, its increasing adoption in standards such as NVMe for smart storage~\cite{ebpf-nvme-standard}, and its security guarantees, it also has major known downsides. These include the difficulty of programming in eBPF %
and documented security vulnerabilities~\cite{ebpf-untrusted,ebpf-safe,privbox}. We choose to implement remote-storage pushdown using eBPF primarily due to its widespread adoption, but our design principles would hold for other privileged execution sandbox frameworks, such as WASM or Privbox~\cite{privbox}.

\section{Supporting In-Memory Data Structures}
\label{sec:rocksdb}

A significant challenge in implementing remote-storage pushdown is utilizing a storage system's in-memory data structures, which are stored at the client, with the on-disk lookup, which occurs remotely. Our solutions to this challenge apply generally to any storage system that needs to use storage pushdown, but in this section, we focus on how we tackled this problem in the integration with RocksDB, as it utilizes a wide variety of continuously-updated in-memory caches (\eg index and data block caches, Bloom filters) common to many systems~\cite{splinterdb,pebblesdb,leveldb,cassandra}. %

\subsection{RocksDB Architecture}

\paragraph{LSM-tree.}
RocksDB is an LSM-tree based storage engine~\cite{o1996log}. To avoid SSD write amplification and wear out, incoming writes to RocksDB are buffered in memory, in a data structure called the MemTable. %
After they fill up, MemTables are asynchronously flushed to disk as sorted-string table (SST) files, which are composed of sorted key-value pairs. SST files are divided into data and metadata blocks. Metadata blocks include index blocks, which map keys to data blocks, and filter blocks, which contain Bloom filters (these are disabled by default in RocksDB, but are supported by \name).

These SST files are organized into hierarchical levels ($L_0$, $L_1$, ..., $L_N$), where the ``upper'' levels store the latest versions of each key-value pair (\eg $L_0$ is ``higher'' than $L_1$) . The immutable MemTables are flushed into $L_0$, which stores files with overlapping key ranges, whereas files in the lower levels ($L_1, ...,L_N$) hold non-overlapping ranges. RocksDB uses a background thread that periodically scans SST files from adjacent levels (\eg $L_2$ and $L_3$), and combines them into a single file, which is written to the lower level (\eg $L_3$). In the process, RocksDB removes deleted and stale versions of keys, freeing up space. It has been shown that to obtain good performance in LSM-trees~\cite{o1996log}, each level contains a multiple of (\eg 10$\times$) the capacity of the previous level, whereby the lowest level ($L_N$) makes up most of the LSM-tree's capacity.

\vspace{-0.5em}\paragraph{Workflow of a read.}
Since more updated versions of key-value pairs sit at the ``upper'' levels of the tree, when a read request arrives, RocksDB searches for the object in the levels of the tree from top to bottom. It first checks if the key-value pair is in RocksDB's MemTables. If not, it checks in $L_0$'s files, followed by $L_1$, etc. For each file, RocksDB first reads the filter blocks to check if the key might be in that file (with a high probability). If so, it checks the index block to find the relevant data block and searches for the key in the data block. These blocks may or may not be cached in memory by RocksDB: typically many (or all) of the index and filter blocks are cached. Therefore, as it is locating the key within a file, \name may consult an in-memory data structure (\eg a cached index block), followed by disk I/O (\eg a data block). This can create an interleaved pattern of in-memory and disk lookups, which is challenging for \name, since it only provides a significant speedup for a \emph{chain} of storage I/Os.

To facilitate lookups, LSM-trees cache the key ranges each file contains. We make a key observation, which we rely on in our design, that this cache can be used to determine all the files whose range contains a particular key, and therefore \emph{might} be needed to be accessed to satisfy a read operation.

\subsection{Integration with \name}

In integrating with RocksDB, we only focus on accelerating the read path, since writes are already heavily optimized and are not in the critical path (all writes are buffered in memory and written sequentially to disk).
Initially, our prototype only accelerates single-key point lookups. In integrating RocksDB with \name we make the following two key contributions: (a) a new technique, \emph{query splitting}, that splits the in-memory accesses from the storage accesses to maximize the effectiveness of storage pushdown, and (b) periodic cache sampling to keep the storage engine's cache fresh. %
We also describe some of the challenges of converting components of a C++ commercial storage engine to eBPF.
Overall, our RocksDB integration with \name required $\sim 700$ LoC in changes to RocksDB itself and $\sim 1,200$ LoC to reimplement RocksDB's parser function in eBPF.

\subsubsection{Query Splitting}

\begin{figure}[!t]
    \centering
    \includegraphics[width=0.9\columnwidth]{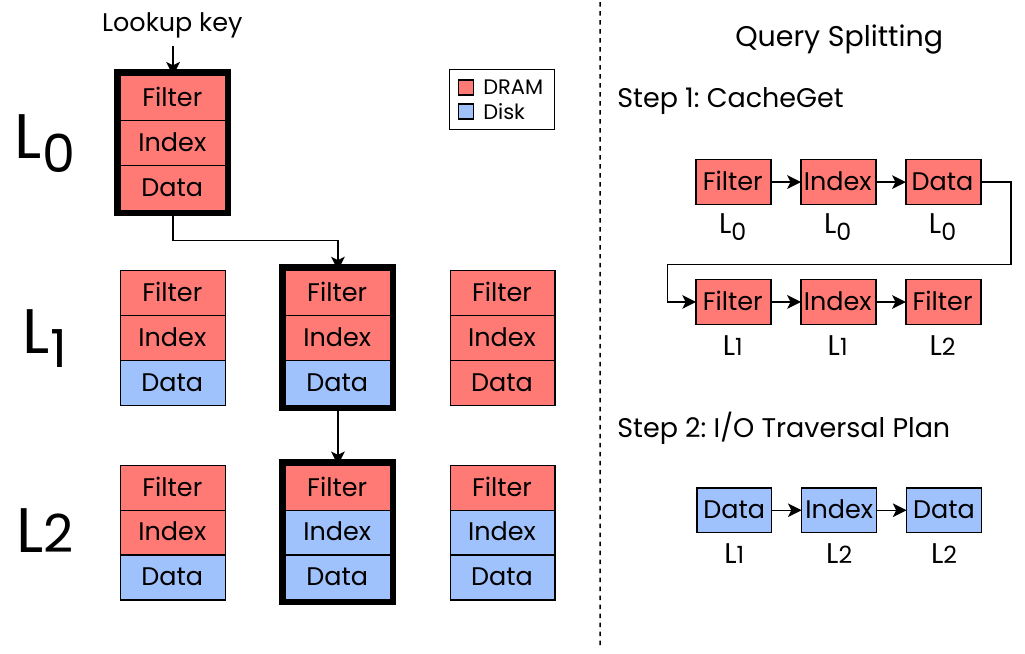}
    \vspace{-1em}
    \caption{RocksDB LSM-tree traversal and query splitting.}
    \label{fig:rocksdb-cacheget}
    \vspace{-1.75em}
\end{figure}

In order to maximize the efficiency of storage lookups, ideally the remotely-executed function would involve a chain of I/O requests that exclusively access the disk, rather than in-memory data structures. To this end, we split the query into two phases: first, the query accesses all the in-memory data structures that \emph{may} be involved in a key lookup, and it then triggers all the required I/O as a series of dependent I/O functions at the remote node. Recall that due to the design of LSM-trees, the host can determine all the files that may be needed to complete the operation. We are guided by the observation that accessing RocksDB's in-memory data structures is orders of magnitude less costly than storage I/O (especially when that I/O is over the network). Therefore, there is little cost in potentially accessing more in-memory data structures than would have been necessary otherwise.

To this end, we implement a new function in RocksDB, \textsc{CacheGet}. \textsc{CacheGet} is based on RocksDB's existing per-SST file \textsc{Get} function, but it only reads data from that file's cached blocks without triggering any storage I/O. Our prototype runs \textsc{CacheGet} on all files that may contain the key, across all levels. Based on the results of this function (\eg whether a key is found in a particular data block), our prototype creates an \emph{I/O traversal plan}, which only fetches uncached blocks that may be required in the I/O traversal. The traversal plan skips files that have both index and data blocks cached, those that are filtered by a per-file Bloom filter, and it starts I/O at the data block if the index block is cached. 

We illustrate this process in Figure~\ref{fig:rocksdb-cacheget}. In this example, three files across $L_0$, $L_1$, and $L_2$ have key ranges that contain the requested key. All of these files may need to be read to return the key. As the figure shows, these files' blocks are partially or fully cached. \textsc{CacheGet} retrieves the blocks in the potential path of the key that are cached in DRAM. The remaining uncached blocks form the on-disk I/O traversal plan. If the key is not found in the cached blocks, our RocksDB prototype issues a \syscall system call, with the files from the I/O traversal plan.
Note that \syscall does not blindly follow the I/O traversal plan to completion, and it terminates once the key is found. For example, if a key exists in $L_1$, the parser will not read data from $L_2$, since the value in $L_1$ will be more up-to-date than other versions lower in the tree. As such, \textsc{CacheGet} will often be called on more files than necessary, since RocksDB does not know the \emph{minimal} set of files that will be involved in the traversal in advance.

\subsubsection{Cache Sampling}
\label{sec:rocksdb:cache-update}
In RocksDB, all data blocks that were accessed as part of a query are cached by default. However, \name returns only the final query result, not the intermediate blocks. With \name, in some cases RocksDB's cache might be ``missing'' intermediate data blocks. This problem would exist in any system with storage pushdown that caches intermediate results. A naive solution would be to return all of the data blocks accessed in the query, but this would negate much of the benefit of storage pushdown, by increasing network bandwidth (and CPU costs) significantly. As such, we seek another way to keep the cache fresh.
We are inspired by prior work on cache sampling~\cite{cachesim}, where a low percentage of the randomly-sampled requests update the cache with recently-accessed intermediate data blocks (as well as with the end result). We implement this idea by sending a small percentage of requests through the slower, non-offloaded ``normal" read path, which caches all the intermediate results. Empirically, we find this technique is quite effective with sampling rates as low as 0.1-1\% (see \S\ref{sec:sampling-rate-eval}).

\subsubsection{Converting RocksDB's I/O Lookups to eBPF}
The final practical obstacle in integrating RocksDB was converting RocksDB's I/O lookups into eBPF functions.
We converted RocksDB's SST file parsers, which parse SST index and data blocks, to eBPF. To the best of our knowledge, our parser is the first example of such complex logic implemented in eBPF. Our eBPF parser first uses the generated I/O traversal plan, which is stored as an array in the scratch buffer, to parse each file at the correct file offset and parsing stage (\ie index or data block). If the requested key is not found in the parsed data, the eBPF parser will resubmit a new I/O request for the next file in the plan. This process continues until either the key is found or all relevant files have been traversed. If the eBPF parser fails, we fall back to the regular RocksDB read path. This occurs infrequently when index blocks are pinned, as such a request will only fail when a RocksDB data block (by default $\sim$4KB) is split between ext4 extents, which tend to be significantly larger than 4KB for (immutable) SST files.

Converting RocksDB's parser to eBPF frequently ran into the limitations of the eBPF programming framework, due to the need to rewrite RocksDB's C++ I/O functions into C. It also required adding various memory access and loop iteration checks to pass verification. Due to eBPF's instruction-count and jump limits, many logical RocksDB functions have to be split up into smaller functions so they can be verified successfully. We did so using \emph{function-by-function verification}~\cite{bpf-fbfv} by splitting the logical functions into “isolated” components that are independently verified. Our eBPF parser demonstrates the viability of implementing complex storage functionality in eBPF, despite its restrictions.

\section{Evaluation}
\label{sec:evaluation}

This section answers the following questions. Questions 1-4 are answered with RocksDB, and Q5 targets the other systems (WiredTiger and BPF-KV):
\begin{denseenum}
\item[{\bf Q1:}] How does \name perform under different network protocols, storage devices, and cache sizes compared to the baseline NVMe-oF and local environments?
\item[{\bf Q2:}] What is \name's impact on CPU and network utilization and energy consumption?
\item[{\bf Q3:}] How do different sampling rates affect performance?
\item[{\bf Q4:}] What is the performance impact of version mismatches?
\item[{\bf Q5:}] How does \name generalize to other storage systems?
\end{denseenum}

\begin{figure*}[!t]
    \centering
    \begin{subfigure}[t]{0.31\textwidth}
         \includegraphics[width=\columnwidth]{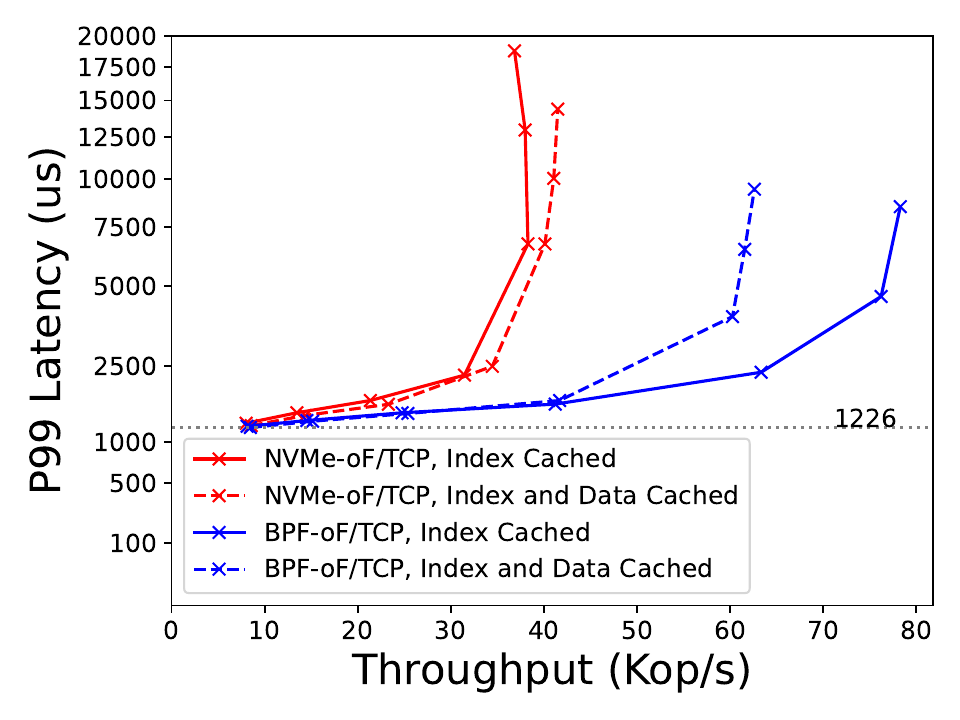}
         \label{fig:rocksdb-nand-thru-lat}
    \end{subfigure}
    \hfill
    \begin{subfigure}[t]{0.34\textwidth}
        \includegraphics[width=\columnwidth]{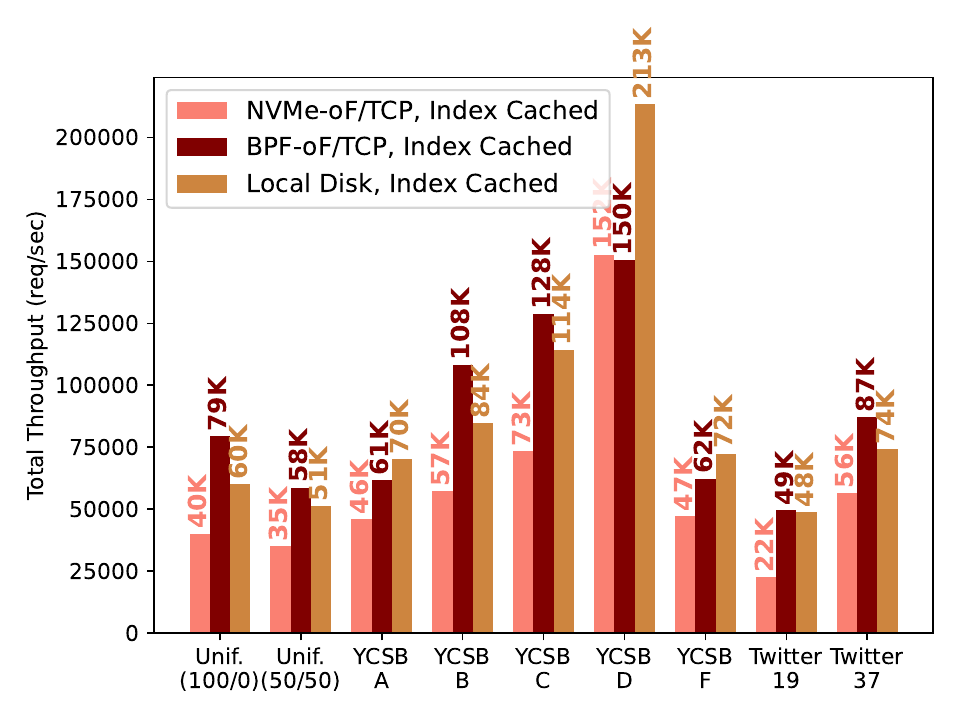}
        \label{fig:index-nand}
    \end{subfigure}
    \hfill
    \begin{subfigure}[t]{0.34\textwidth}
        \includegraphics[width=\columnwidth]{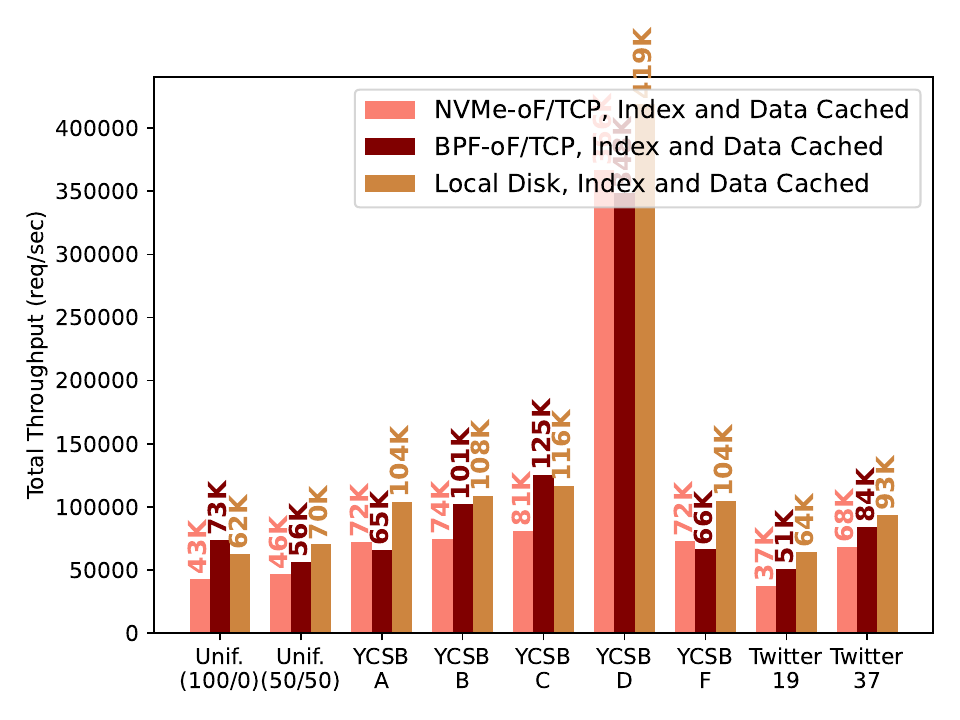}
        \label{fig:data-nand}
    \end{subfigure}
    \vspace{-2em}
    \caption{RocksDB (a) uniform read throughput-latency and throughput (b) without and (c) with data blocks cached of \name vs. NVMe/TCP on NAND SSD.}
    \label{fig:rocksdb-NAND}
    \vspace{-1.5em}
\end{figure*}

\subsection{Experimental Setup}
\label{sec:experimental-setup}

We conduct our experiments in three different CloudLab storage configurations~\cite{cloudlab}, using NAND and Optane SSDs. %
We measure an average roundtrip latency of 30\us and 18\us for TCP and RDMA, respectively.

\vspace{-0.5em}
\paragraph{NAND configuration.}
This is a typical enterprise NAND SSD setup. Both host and target are c6525-100g machines in CloudLab. Each machine has a 24-core AMD EPYC 7402P CPU, 128~GB of RAM, two 1.6~TB NAND SSDs (Dell Ent NVMe AGN MU U.2), and a Dual-port Mellanox ConnectX-5 NIC with a 100~Gbps network. %
The NAND SSD's read latency is 90\us and can sustain 700K read IOPS.

\vspace{-0.5em}
\paragraph{Optane configuration.} This setup uses a low-latency storage device. The host is a c6525-100g machine, while the target is d750 with Intel Optane SSD. The target has two 16-core Intel Xeon Gold 6326 CPUs, 128~GB of RAM, a 400~GB Optane 5800X SSD, and a Quad-port BCM57504 NetXtreme-E on a 25~Gbps network. %
The Optane SSD has a 3\us 512B read latency and can sustain 5M read IOPS.

\vspace{-0.5em}
\paragraph{RDMA offloading configuration.} This setup uses RDMA NIC offloading. Both host and target are r6525, with 32-core AMD EPYC 7543 CPUs, 256~GB of RAM, one 1.6~TB NAND SSD (Dell Ent NVMe AGN MU U.2), and a Dual-port Mellanox ConnectX-6 100~GB NIC on a 100~Gbps link.

\vspace{-0.5em}\paragraph{Software configuration.} To measure the throughput per core, we increase the number of threads per core up until saturation, and use CPU pinning to bind each thread to a specific core, and flow-steering to associate each flow with a specific host and target core. To measure energy, we use AMD's \textmu{}Prof tool. To make our results reproducible, we disable hyperthreading and address space randomization. We run Ubuntu 20.04 and Linux 5.12.0. %

\vspace{-0.5em}\paragraph{Evaluated systems.}
We implement \name on three systems: RocksDB, WiredTiger and BPF-KV.
RocksDB is a complex and popular key-value store, %
and we test it with \name to show whether a pushdown approach would benefit state-of-the-art storage applications. %
WiredTiger is a very simple production-grade key-value store used primarily by MongoDB. %
BPF-KV is a bare-bones academic key-value store custom-designed for running eBPF functions~\cite{xrp}.
All systems bypass the OS page cache with direct I/O.

\vspace{-0.5em}\paragraph{RocksDB setup.}
We build upon RocksDB 7.7.3 with a 100~GB database in 5 layers ($L_0-L_4$) and 3 cores (which saturate our enterprise SSD); we use 24 and 20 threads-per-core for NAND and Optane SSD, respectively. Unless noted, all index blocks are cached and Bloom filters are disabled (RocksDB's default). We run RocksDB with both TCP and RDMA; our results focus on TCP, as it is more prevalent. By default, we use a cache sampling rate of 1\% (\ie a random 1\% of reads go through the regular non-\name read path).

\vspace{-0.5em}\paragraph{Workloads.}
To test \name, we run the YCSB~\cite{ycsb} benchmark suite under Zipfian (0.99) key distributions, along with uniform read and read-write workloads. We do not run YCSB E, because our RocksDB integration does not yet accelerate scans. We also test \name with a production trace taken from clusters 19 and 37 of the Twitter cache workloads~\cite{twittertraces}. These traces are read-dominant (75\% and 63\%, respectively) and have low skew ($\alpha = 0.7$ and $0.4$, respectively), ensuring that their working set does not fit entirely in memory.

\subsection{\name With Different Setups (Q1)}
\paragraph{TCP with NAND SSD.}
\label{sec:eval-tcp-nand}

Figure~\ref{fig:rocksdb-NAND} shows RocksDB's throughput with \name with NAND and TCP, compared to NVMe/TCP and to local RocksDB (not over the network).\footnote{The TCP results in Figure~\ref{fig:rocksdb-local-vs-nvmeof-tcp} were obtained on the offloading machine in order to compare them to RDMA with offloading. As such, the TCP results in Figure~\ref{fig:rocksdb-NAND} are not identical to those in Figure~\ref{fig:rocksdb-local-vs-nvmeof-tcp}.}%
Without a data cache, \name can achieve 1.4-2.2$\times$ higher throughput than NVMe/TCP, and in many cases (including the Twitter cache traces, which are real-world workloads) matches or exceeds the throughput of running on local disk. This is because \name eliminates repeated traversals of the kernel I/O stack on the target, which in some cases outweighs the added network processing cost.
This trend continues when adding a data cache, where \name provides 1.2-1.7$\times$ higher throughput than NVMe/TCP.

The relative speedup of \name with a data cache is generally lower, because some lookups may be cached, thereby reducing the average number of roundtrips saved by \name.
The only cases where \name falls slightly short are YCSB A, D, and F, with data blocks cached. Since these workloads are highly-cacheable, they issue very few I/Os per request, and \name's slightly higher overhead for in-memory requests (\eg its preemptive access of all possible in-memory caches in each query) leads to a slight slowdown. However, with just index blocks cached, \name provides a speedup in all workloads except YCSB D, and it even beats the baseline with cache for a majority of the workloads.
Figure~\ref{fig:rocksdb-NAND} also shows that \name reduces tail latency by up to 2$\times$. %

\begin{figure*}[t]
    \centering
    \begin{subfigure}[t]{0.245\textwidth}
        \includegraphics[width=\columnwidth]{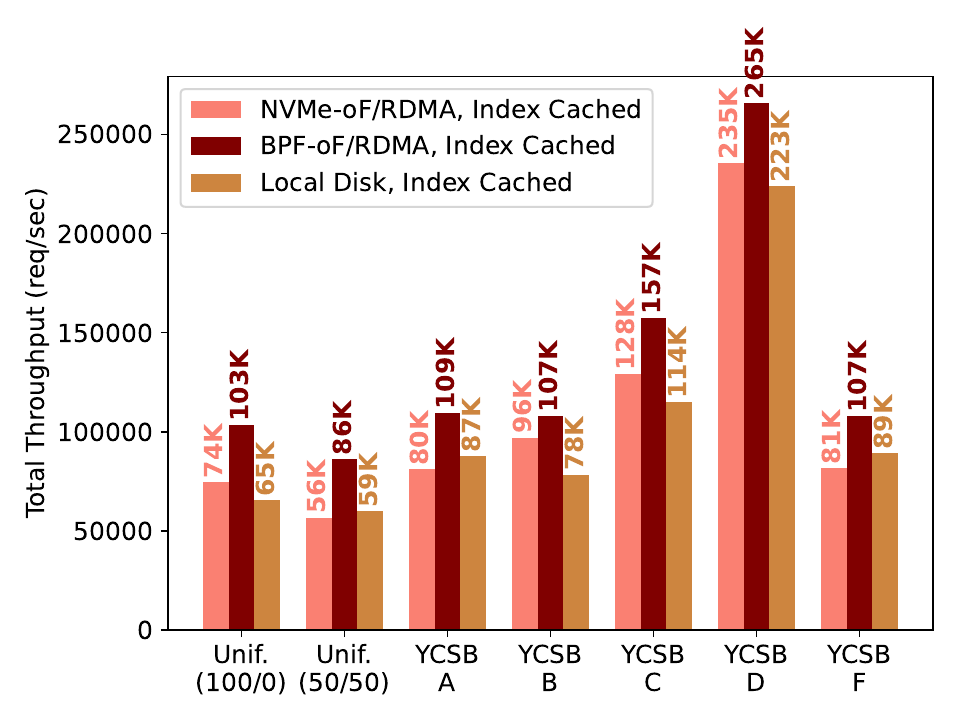}
        \caption{NVMe/RDMA with NAND SSD without data blocks cached.}
        \vspace{-1em}
        \label{fig:index-nand-rdma}
    \end{subfigure}
    \hfill
    \begin{subfigure}[t]{0.245\textwidth}
        \includegraphics[width=\columnwidth]{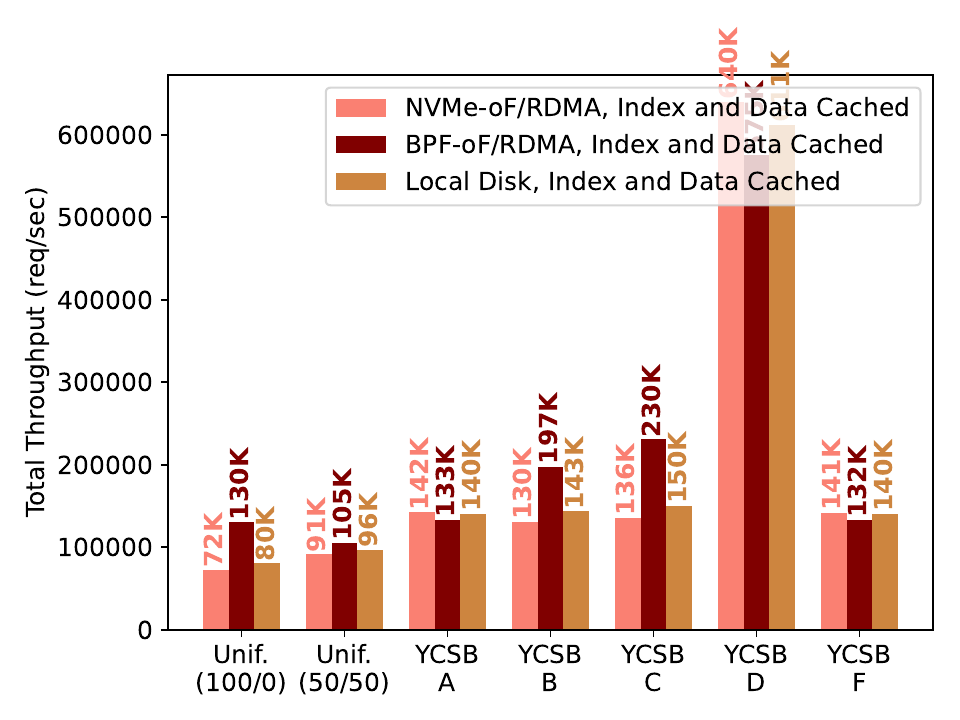}
        \caption{NVMe/RDMA with NAND SSD with data blocks cached.}
        \label{fig:data-nand-rdma}
    \end{subfigure}
    \hfill
    \begin{subfigure}[t]{0.245\textwidth}
        \includegraphics[width=\columnwidth]{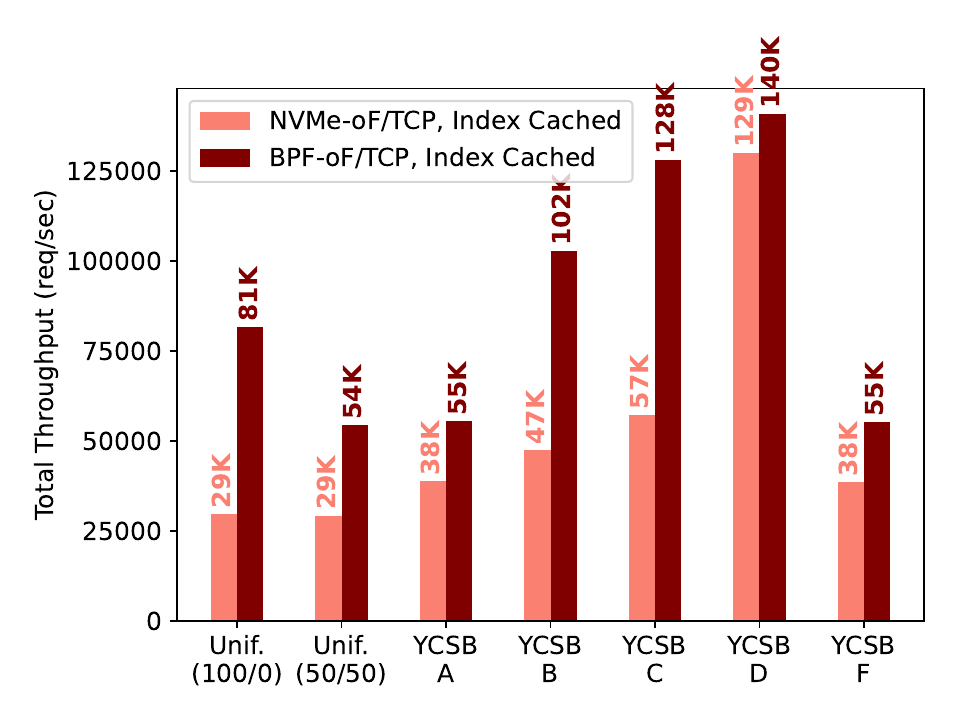}
        \caption{NVMe/TCP with Optane SSD without data blocks cached.}
        \label{fig:index-optane}
    \end{subfigure}
    \hfill
    \begin{subfigure}[t]{0.245\textwidth}
        \includegraphics[width=\columnwidth]{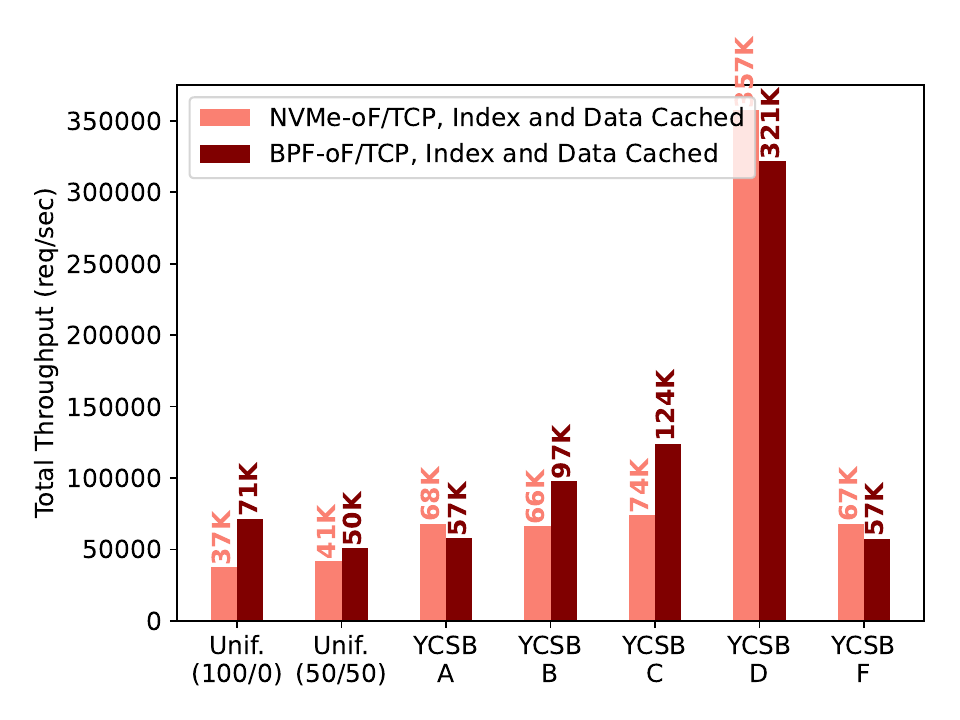}
        \caption{NVMe/TCP with Optane SSD with data blocks cached.}
        \label{fig:data-optane}
    \end{subfigure}
    \vspace{-0.5em}
    \caption{RocksDB throughput under different configurations.}
    \label{fig:rocksdb-NAND-rdma}
    \vspace{-1.4em}
\end{figure*}

\vspace{-0.5em}\paragraph{RDMA with NAND SSD.}
Figures~\ref{fig:index-nand-rdma} and \ref{fig:data-nand-rdma} show the throughput of RocksDB with \name with NAND and RDMA, compared to \emph{NIC offloaded} NVMe/RDMA and to local RocksDB. Even though \name does not take advantage of offloading the RDMA requests to the NIC, \name is up to 1.8$\times$ faster than offloaded NVMe/RDMA. Since RDMA is very CPU-efficient it achieves an up to 1.6$\times$ speedup even comapred to the local disk baseline. In our experiments, there was no noticeable difference in the baseline's throughput between offloaded and non-offloaded RDMA. The reason for this is that while offloading the RDMA protocol to the NIC saves CPU at the target, the target's CPU is usually not a bottleneck with RocksDB.

\vspace{-0.5em}\paragraph{TCP with Optane SSD.}
We also test \name with Optane, providing a glimpse of \name's benefits as SSDs become faster. \name provides even greater speedups with Optane (Figures~\ref{fig:index-optane} and \ref{fig:data-optane}), providing 2-2.8$\times$ higher throughput. This is because Optane can support a high I/O bandwidth, while in the NAND SSD experiments I/O bandwidth may become a bottleneck. \name also provides a consistent improvement (up to 2.6$\times$) in tail latency in all workloads. In addition, we see a surprising result: \name without a cache yields higher throughput than \name with a cache. We discuss this behavior below.

\vspace{-0.5em}\paragraph{Caching and Storage Pushdown.}
Caching has an interesting and unexpected interaction with \name. In certain cases, running RocksDB with a cache can actually slow down the system, as seen in Figure~\ref{fig:data-optane}. In the baseline NVME-oF case, RocksDB transfers all the intermediate blocks over the network until it finds the final result. Thus, maintaining a cache for these blocks incurs little additional cost. However, when using storage pushdown with \name, RocksDB only transfers the final result over the network, and we employ cache sampling to keep the cache fresh with intermediate results. %
In some workloads, the cost of keeping the cache fresh outweighs its benefits. This balance depends on the workload type, the cache miss rate, and the I/O cost.

\vspace{-0.5em}\paragraph{Non-Default Configurations.}
By default, we evaluated RocksDB with indices pinned and without Bloom filters, as these are the default configuration options in RocksDB. For completeness, we also tested (1) with Bloom filters enabled and (2) with indices not pinned. The results were not very different than with our default configurations: \name provides significant speedups in both cases, with up to 2$\times$ improvement with Bloom filters and 1.8$\times$ improvement with unpinned indices (for more details, see \S\ref{sec:appendix-rocksdb-non-default}).

\subsection{CPU, Network, and Energy Savings (Q2)}
\name provides significant CPU, network, and energy (and carbon emissions) savings, which are directly correlated with performance increases. To measure these savings, we monitor the CPU time spent \emph{on both the host and target}, network bytes exchanged, and energy spent while running a benchmark. We present these results (normalized per-request) for uniform reads with a 1~GB data cache, in Table~\ref{tab:rocksdb-cpu-network}. For TCP, \name consumes 37\% fewer CPU cycles (measured in mCPUs, \ie $1/1000$th of a CPU) and generates 23\% less network traffic, resulting in a roughly 70\% throughput improvement. It does so while consuming 36\% less energy per request. In the offloaded RDMA setting, \name consumes 29\% and 19\% less CPU and network traffic per request, respectively, resulting in a 65\% throughput improvement. Energy measurements were not available in the RDMA machines. Additionally, \name should can yield even lower carbon emissions due to the decreased reliance on energy-hungry DRAM.

\begin{table}[h]
    \vspace{-0.5em}
    \centering
    \caption{Resource consumption with uniform read and 1~GB data cache.}
    \vspace{-0.5em}
    \label{tab:rocksdb-cpu-network}
    \resizebox{\columnwidth}{!}{
        \begin{tabular}{|l|c|c|c|c|}
        \hline
        \textbf{Setup} & \textbf{Throughput}  & \textbf{CPU}         & \textbf{Network}    &\textbf{Energy}   \\
                       & \textbf{(req/s)}     &  \textbf{(mCPU / req)} & \textbf{(kB / req)}&\textbf{(mW / req)} \\ \hline
        NVMe-oF/TCP              & 48279                      & 0.10                   & 12.07                   & 3.63                 \\ \hline
        \name/TCP               & 69935                      & 0.06                   & 9.27                    & 2.33                 \\ \hline
        NVMe-oF/RDMA (offload)   & 77553                      & 0.04                    & 12.3                    & -                    \\ \hline
        \name/RDMA              & 120991                     & 0.03                    & 10.02                   & -                    \\ \hline
        \end{tabular}
    }
    \vspace{-1.5em}
\end{table}

\subsection{Sampling Rate (Q3)}
\label{sec:sampling-rate-eval}
\begin{figure}[t]
    \centering
    \includegraphics[width=0.875\columnwidth]{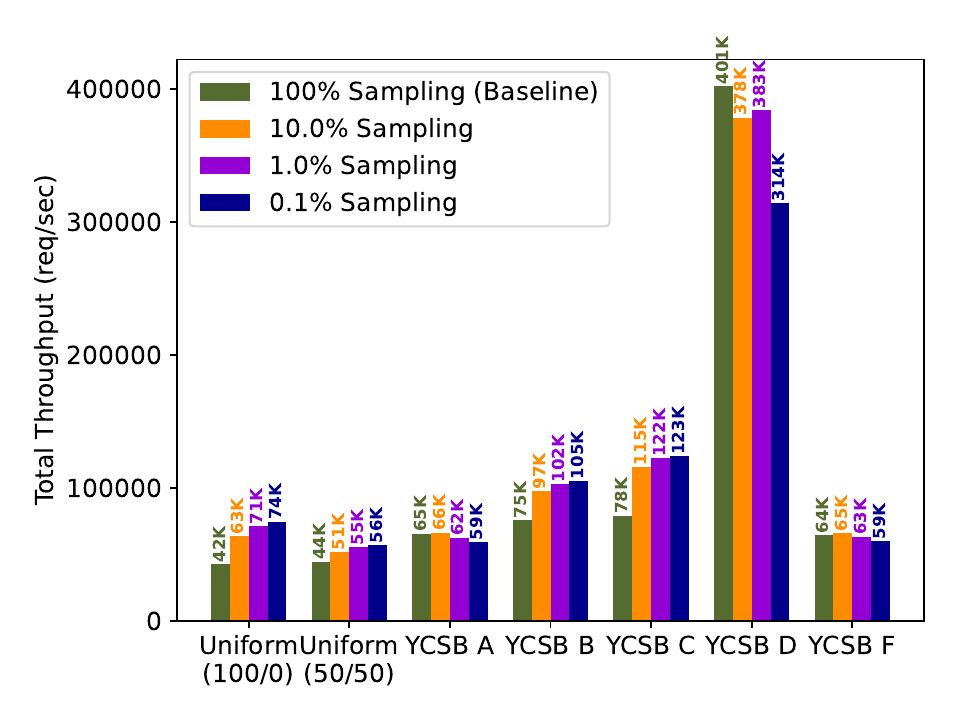}
    \vspace{-1em}
    \caption{RocksDB throughput with different sampling rates with cache under \name using TCP and NAND SSD.}
    \label{fig:rocksdb-sampling}
    \vspace{-1.5em}
\end{figure}

As discussed in \S\ref{sec:rocksdb:cache-update}, our RocksDB integration sends a small portion of reads through the normal read path in order to keep the cache warm. We evaluate the impact of this sampling rate on RocksDB's performance. Figure~\ref{fig:rocksdb-sampling} shows that throughput peaks at a sample rate between 0.1\% and 10\%, except for YCSB D, which is highly cacheable and thus benefits from a higher sampling rate. Similarly, we see a slight decrease in performance in YCSB A and F at higher sampling rates. These workloads are write-heavy and highly skewed, which means that as the popular keys are written repeatedly, they reappear in the higher levels of the LSM tree and shorten the resubmission chains, leading to a smaller performance improvement. Based on these experiments, we run \name with a 1\% sampling rate by default.

\subsection{Version Mismatches (Q4)}
As discussed in \S\ref{sec:versioning}, \name versions file extent metadata in order to guarantee safe concurrent access to files for the offloaded applications. A version mismatch occurs when the synchronized file system metadata in the target becomes stale. A version mismatch results in a failed request, which is then retried through the non-accelerated read-path. Thus, version mismatches are very costly: they approximately decrease throughput by 1 request and more than double the requests' latency. Fortunately, they are very rare. Across all benchmarks, we observed only 20 mismatches, out of millions of requests per benchmark. The only benchmark that had a slightly higher number of mismatches was YCSB D, due to its write-heaviness, but they still occurred in only 0.03\% of requests. Thus, the performance impact of version mismatches is negligible in our integration with RocksDB.

\subsection{WiredTiger (Q5)}
\label{sec:eval-wiredtiger}

\begin{figure*}[t]
    \centering
    \begin{subfigure}[t]{0.32\textwidth}
        \includegraphics[width=\columnwidth]{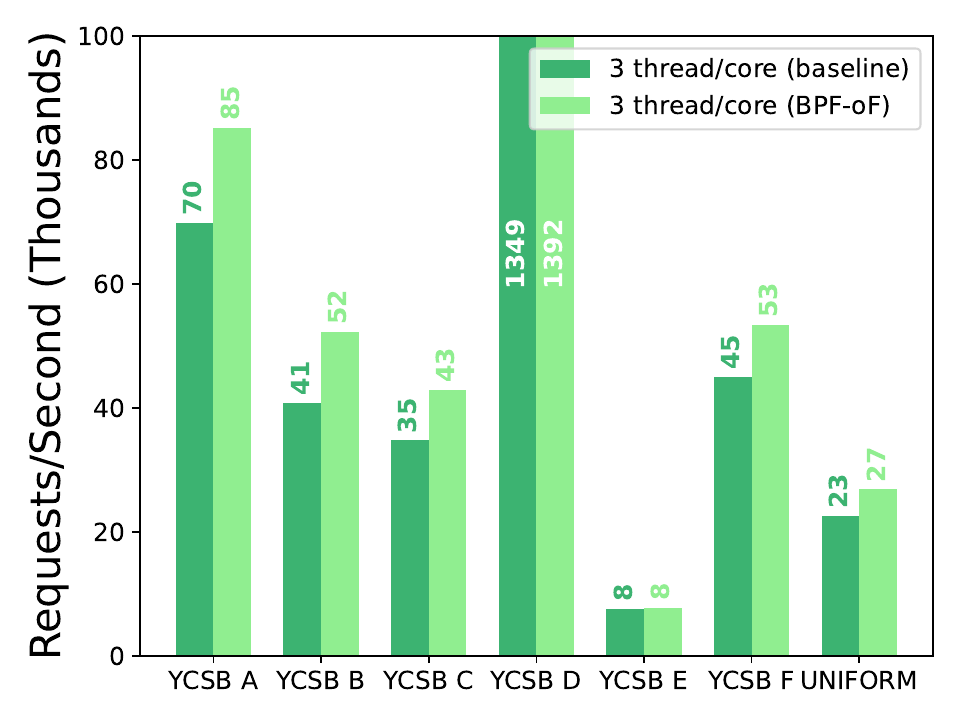}
        \caption{Throughput with WiredTiger.}
        \label{fig:wiredtiger_nand}
    \end{subfigure}
    \hfill
    \begin{subfigure}[t]{0.32\textwidth}
        \includegraphics[width=\columnwidth]{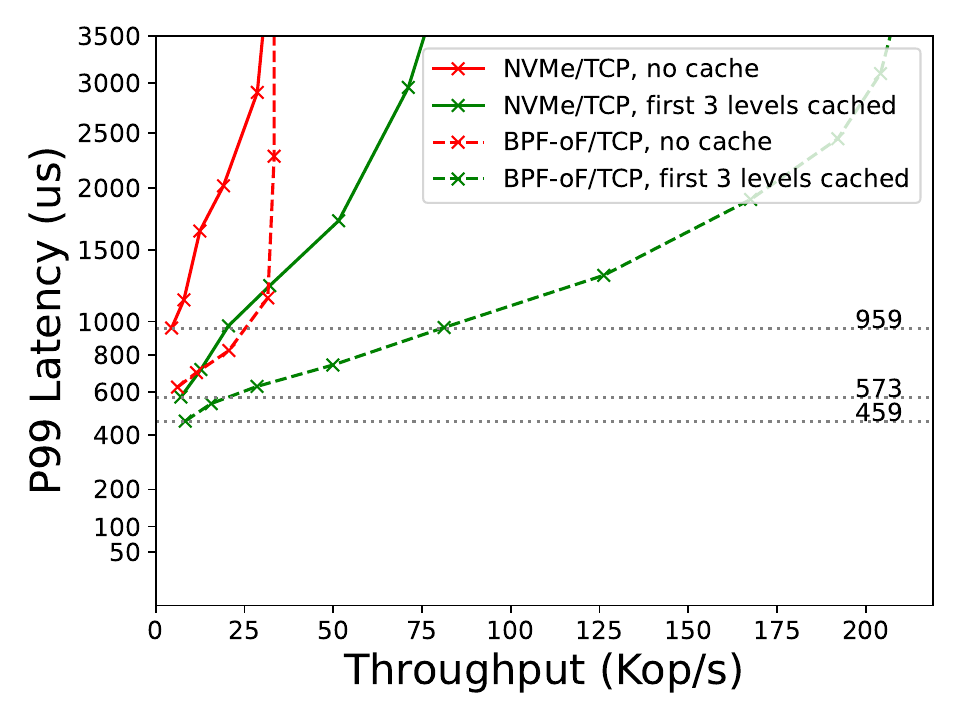}
        \caption{Throughput-latency of BPF-KV with random reads.}
        \label{fig:bpfkv-3core-TCP-NAND}
    \end{subfigure}
    \hfill
    \begin{subfigure}[t]{0.32\textwidth}
        \includegraphics[width=\columnwidth]{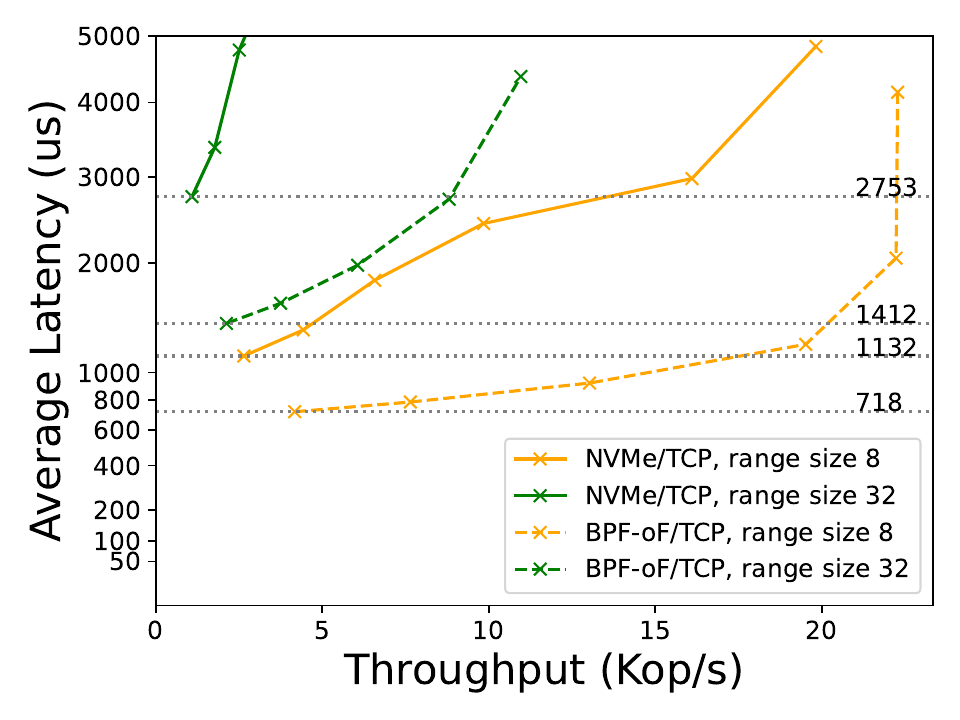}
        \caption{Random range queries with BPF-KV.}
        \label{fig:bpf-kv-range}
    \end{subfigure}
    \vspace{-0.5em}
    \caption{WiredTiger and BPF-KV evaluations with TCP and NAND SSD.}
    \label{fig:wt-bpf-kv-eval}
    \vspace{-1.5em}
\end{figure*}

We run WiredTiger in its LSM-tree version, wherein data is split into levels, and each level is stored in a single file. Each file uses a B-tree index where the data is stored as the leaf nodes. %
We configure the B-tree page size to be equal to 512B. The database contains 1G items with a 16B object size.
We test WiredTiger with \name using the YCSB benchmark on 4 cores, the minimal number of cores to eliminate CPU contention between foreground and background threads. %
Figure~\ref{fig:wiredtiger_nand} shows that for read-write workloads such as YCSB A, \name has 20\% higher throughput. For read-heavy workloads such as YCSB B, YCSB C and uniform, the improvement reaches 30\% when using an Optane SSD, as seen in Figure~\ref{fig:wt-optane} in the Appendix.
The performance gain of WiredTiger is more modest than that of RocksDB (and BPF-KV), since WiredTiger issues far fewer I/Os, and therefore reducing the time (and CPU consumption) of I/O leads to a more limited speedup.

\subsection{BPF-KV (Q5)}
\label{sec:bpf-kv}

BPF-KV~\cite{xrp} is a toy high-performance key-value store optimized for eBPF functions that was introduced by the XRP paper. It organizes its data in a B$^{+}$-tree, can cache the top-most levels of the tree, and uses fixed-sized 512B key-value pairs. The leaves of the tree contain a pointer to the values, which are stored in an unsorted log. We use a 67~GB database file that has 6 index levels and 1 level for the log.
We conducted a full suite of experiments with BPF-KV to evaluate \name's performance under different network protocols, storage devices, and caching strategies. We present a subset of the results here, with further results presented in \S\ref{sec:appendix-bpf-kv}.

\vspace{-0.5em}\paragraph{Point queries.}
We run a simple microbenchmark using a BPF-KV instance containing 8B keys with 64B values stored on the NVMe-oF target.
Figure~\ref{fig:bpfkv-3core-TCP-NAND} compares the results of running \name vs. regular NVMe-oF on NAND SSDs on three cores.
\name consistently provides about 2.6$\times$ higher throughput than NVMe/TCP when the upper levels of BPF-KV are cached. Note that when no cache is used, the access pattern limits the SSD to 120K IOPS (which translates to 30K requests per second at 7 resubmissions per request), because the upper level nodes become read hot-spots in the NVMe queues. Note that for NVMe/TCP on Optane, \name provides more than 8$\times$ higher throughput when BPF-KV is run without a cache.
This speedup is due to \name saving significant CPU cycles in processing TCP packets, which can instead be used to handle additional I/O requests.

\vspace{-0.5em}\paragraph{Range queries.}
\name can support more complex operations than single-key lookups. Figure~\ref{fig:bpf-kv-range} shows the results over TCP on NAND with range queries, where a range of keys is fetched sequentially from the leaves of BPF-KV's B$^{+}$-tree. \name can achieve over 5$\times$ higher throughput.

\subsection{Takeaways}
\label{sec:takeaways}

We find that \name can provide significant performance benefits for disaggregated storage systems, of up to 2.8$\times$ higher throughput and 2.6$\times$ lower tail latency with RocksDB. In particular, \name benefits workloads that are read-heavy, and whose working set does not fit in the cache, as these necessitate longer resubmission chains which \name is able to accelerate. %
\name also provides significant CPU and network savings, which are directly correlated with performance increases. Notably, we find that caching with storage pushdown is a mixed bag. When workloads are highly skewed a cache avoids issuing unnecessary network traffic, but in some workloads maintaining a cache can be detrimental due to the high CPU and network costs of regularly updating it. %

\section{Related Work}
\label{sec:related}

Decades of prior work have explored pushing functions close to data~\cite{xrp,kourtis2020safe,lambda-io,wu2021bpf,splinter,kayak,hellerstein1993predicate,levy1994query,adaptive-placement,flexpushdown,extfuse}.
Most of these pushdown mechanisms are application-specific, \ie they allow clients of a particular system (\eg a database) to pushdown specific functions (\eg SQL filters). %

\vspace{-0.5em}\paragraph{Systems that use eBPF.}
Several recent systems utilize eBPF to push computation closer to storage.
Kourtis ~\etal ~propose using eBPF to safely execute user-defined functions on a remote NVM storage server~\cite{kourtis2020safe}, by running functions in userspace at the remote server, without integrating with a networked storage protocol. %
Their paper presents preliminary work and does not provide a full system design, implementation, and evaluation.
BMC~\cite{bmc} implements an eBPF-based cache at a remote server's kernel to service clients' memcached requests to minimize kernel-userspace crossings at the server. Unlike \name, BMC does not support custom user functions at the server.
XRP~\cite{xrp} and ExtFUSE~\cite{extfuse} allow users to bypass some of the kernel and file system's codepath, thus reducing overhead and eliminating kernel-userspace context switches by running user-defined eBPF functions. Similarly, $\lambda$-IO~\cite{lambda-io} provides a BPF framework to execute custom functions within a computational storage device. These systems do not support a disaggregated storage use case. Finally, Electrode~\cite{electrode} uses eBPF extensions in the kernel to accelerate distributed protocols like Multi-Paxos. Electrode's acceleration of coordination protocols is complementary to \name.

\vspace{-0.5em}\paragraph{Other frameworks.}
Splinter~\cite{splinter}, ASFP~\cite{adaptive-placement}, and Kayak~\cite{kayak} are in-memory key-value stores that allow users to ship lightweight Rust functions that are executed as co-routines on a remote in-memory server. %
Redis~\cite{redis-functions} provides a similar capability with Lua. Unlike \name, these systems implement pushdown in the context of a specific system (an in-memory key-value store), and are not a general-purpose networked storage protocol. In addition, since they operate in an in-memory setting they do not contend with the issues that arise in storage (\eg file metadata synchronization).
Some storage systems provide application-specific pushdowns. Amazon S3 Select~\cite{amazon-select} allows users of Amazon S3 to apply specific SQL filters to data when it is fetched from S3 to save network bandwidth. %
Ceph~\cite{lefevre2020skyhookdm} provides similar capabilities with Ceph ``extensions''. These systems also do not provide a general-purpose storage pushdown capability that can be plugged into any storage system.

\section{Conclusion}

This work explores how to design a networked storage protocol that allows the client to push user-defined storage functions to the server over the network, thereby significantly improving both client and server CPU efficiency. We believe there are significant future research challenges in this area, such as adding support for commonly-needed storage capabilities, such as disk arrays, RAID, and encrypted storage. Another promising direction is to allow the functions to fully execute on a smartNIC. %
Furthermore, our insights into the interaction of in-memory and fast remote disks, indicate the potential for designing efficient nearly memory-less disaggregated storage systems.

\label{lastpage}

\section{Acknowledgments}
We would like to thank Tanvir Khan and Haoyu Li for their feedback.
We also thank the CloudLab team for their help in supporting our experiments.
This work was supported by gifts from IBM and Intel,
and by NSF award CNS-2143868.
Tal Zussman was supported by NSF award DGE-2036197.
Ioannis Zarkadas is an Onassis Foundation scholar.

\bibliographystyle{plain}
\bibliography{database}

\clearpage
\section{Appendix}
\label{sec:appendix}

\subsection{RocksDB Non-Default Configurations}
\label{sec:appendix-rocksdb-non-default}

\begin{figure}[t!]
    \centering
    \begin{subfigure}[t]{0.40\textwidth}
        \includegraphics[width=\columnwidth]{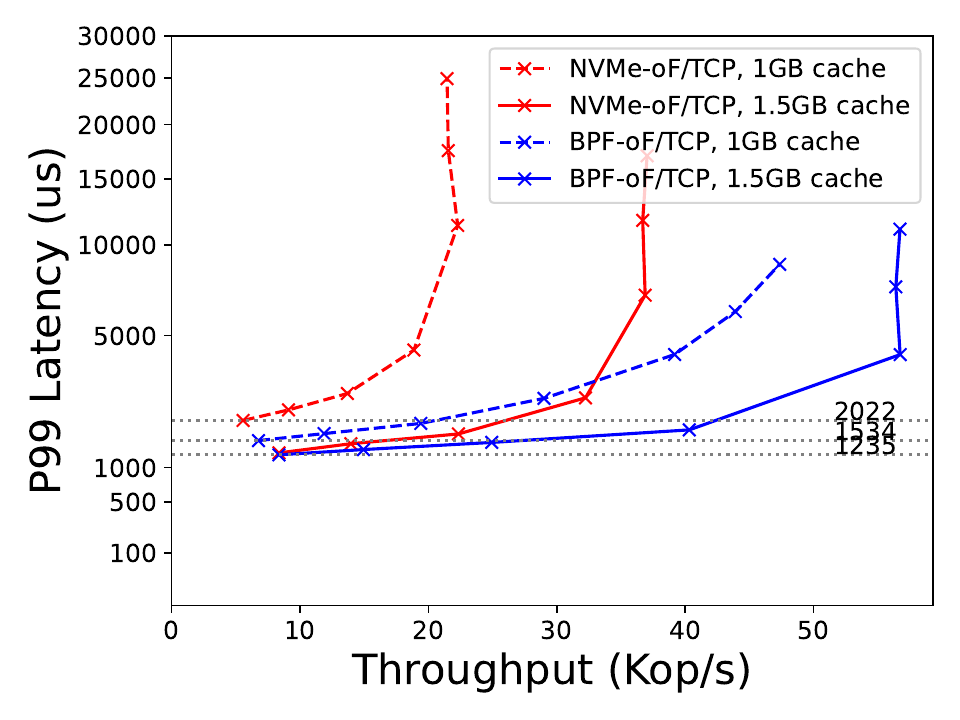}
        \vspace{-2em}
        \caption{Indices unpinned.}
        \label{fig:rocksdb-index-unpinned-thru-lat}
    \end{subfigure}
    \begin{subfigure}[t]{0.40\textwidth}
        \includegraphics[width=\columnwidth]{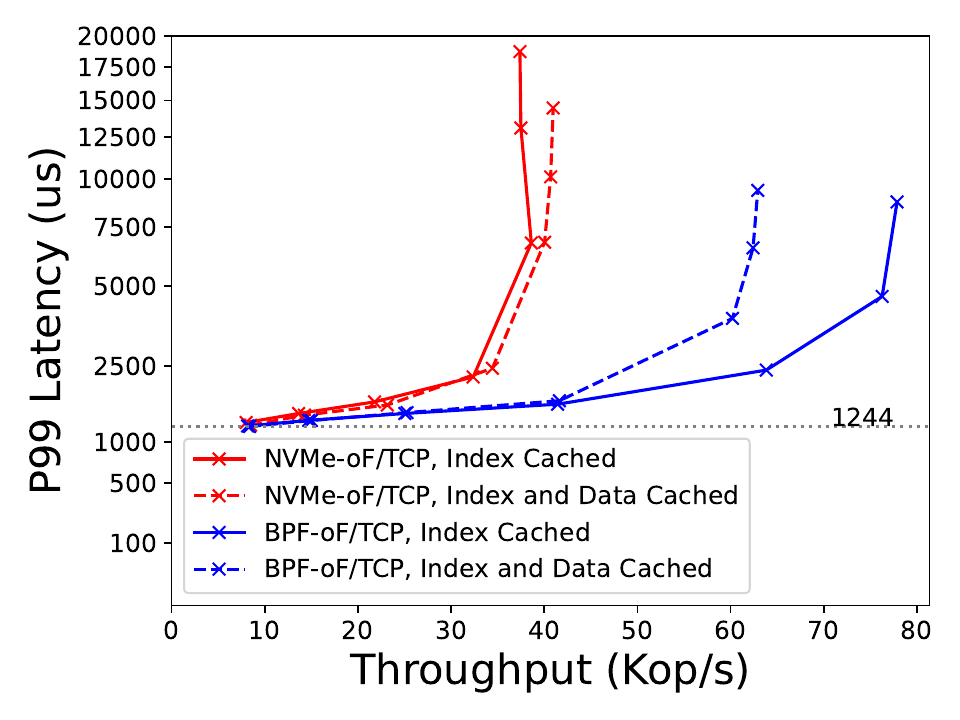}
        \vspace{-2em}
        \caption{Bloom filters enabled.}
        \label{fig:rocksdb-bloom-thru-lat}
    \end{subfigure}
    \vspace{-0.5em}
    \caption{Tail latency and throughput of \name vs. NVMe-oF/TCP on RocksDB with NAND SSD for a uniform read workload for non-default configurations.}
    \label{fig:rocksdb-non-default}
    \vspace{-2em}
\end{figure}

When running with indices not pinned, we did not run a no-cache configuration, since RocksDB is not designed to operate with neither index blocks nor data blocks cached, leading to very low throughput rates. We also run a configuration with a 1.5~GB cache, as RocksDB must cache both index and data blocks in its block cache. This minor increase allows for increased caching while also ensuring that some index blocks must be read from disk. Figure~\ref{fig:rocksdb-index-unpinned-thru-lat} compares these workloads across \name and NVMe-oF. As expected, the 1.5~GB configuration scales better than the 1~GB configuration. \name leads to a 1.8$\times$ throughput improvement while maintaining lower tail latency.

To test performance with Bloom filters, we ran with both a 1~GB cache and no cache. We did not run with filter blocks not pinned, as RocksDB's filter blocks do not provide a meaningful reduction in disk accesses if they are not stored in memory. Interestingly, enabling Bloom filters did not lead to a significant performance benefit when compared to the default configuration, whether using \name or NVMe-oF. However, \name maintained its performance improvement over NVMe-oF in this configuration as well.

\subsection{WiredTiger Evaluation}
\label{sec:appendix-wiredtiger}

\begin{figure}[th]
    \centering
    \includegraphics[width=0.9\columnwidth]{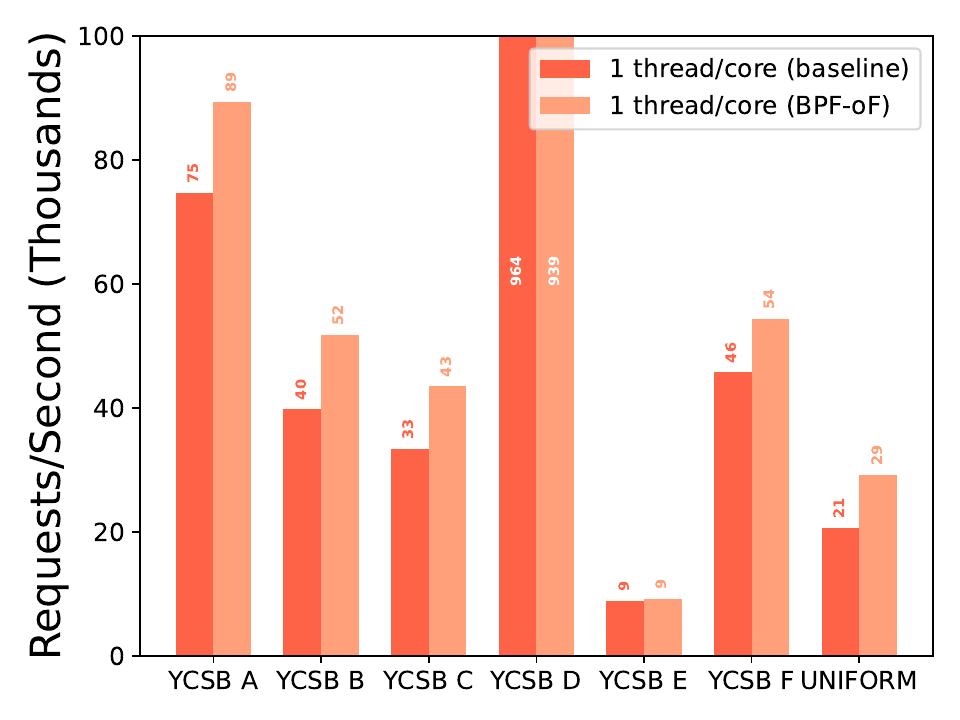}
    \vspace{-1em}
    \caption{Throughput of WiredTiger with TCP and Optane SSD.}
    \label{fig:wt-optane}
    \vspace{-1em}
\end{figure}

We present further WiredTiger results using an Optane SSD and four cores in Figure~\ref{fig:wt-optane}.

\subsection{BPF-KV Evaluation}
\label{sec:appendix-bpf-kv}

\begin{figure}[t!]
    \centering
    \begin{subfigure}{0.4\textwidth}
        \includegraphics[width=\columnwidth]{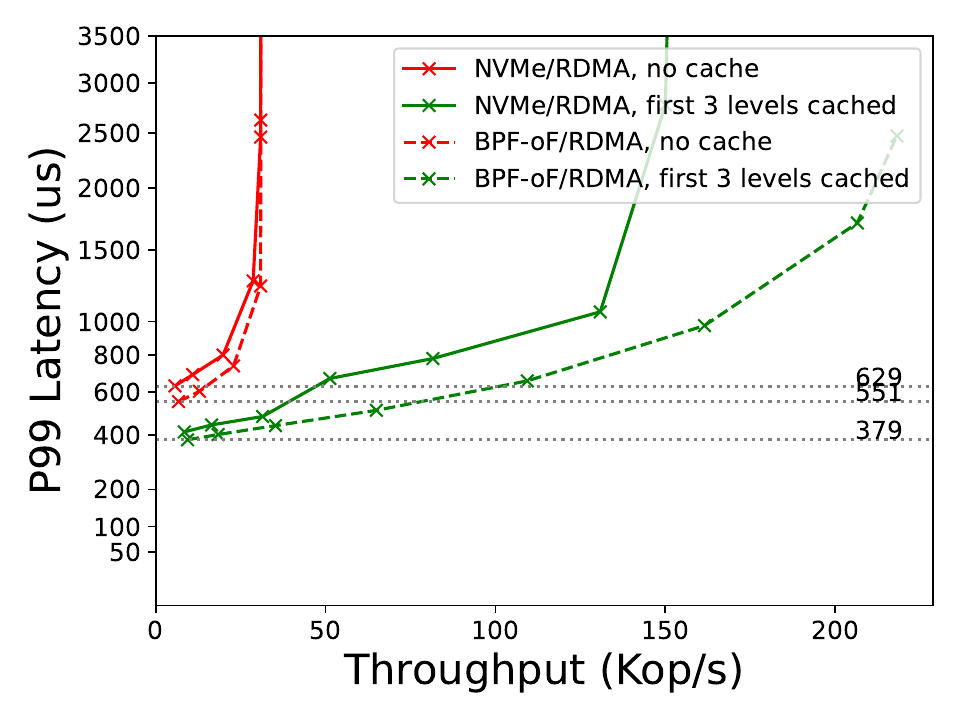}
        \vspace{-2em}
        \caption{3 cores with RDMA, NAND SSD.}
        \label{fig:3core-RDMA-NAND}
    \end{subfigure}
    \hfill
    \begin{subfigure}{0.4\textwidth}
        \includegraphics[width=\columnwidth]{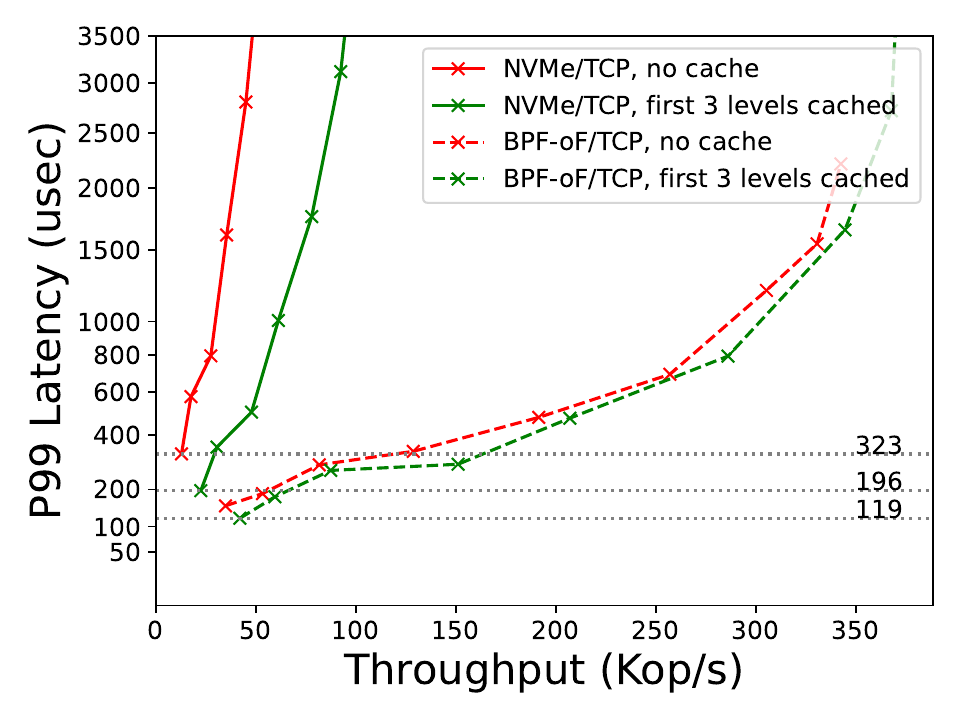}
        \vspace{-2em}
        \caption{3 cores with TCP, Optane SSD.}
        \label{fig:3core-TCP-Optane}
    \end{subfigure}
    \hfill
    \begin{subfigure}{0.4\textwidth}
        \includegraphics[width=\columnwidth]{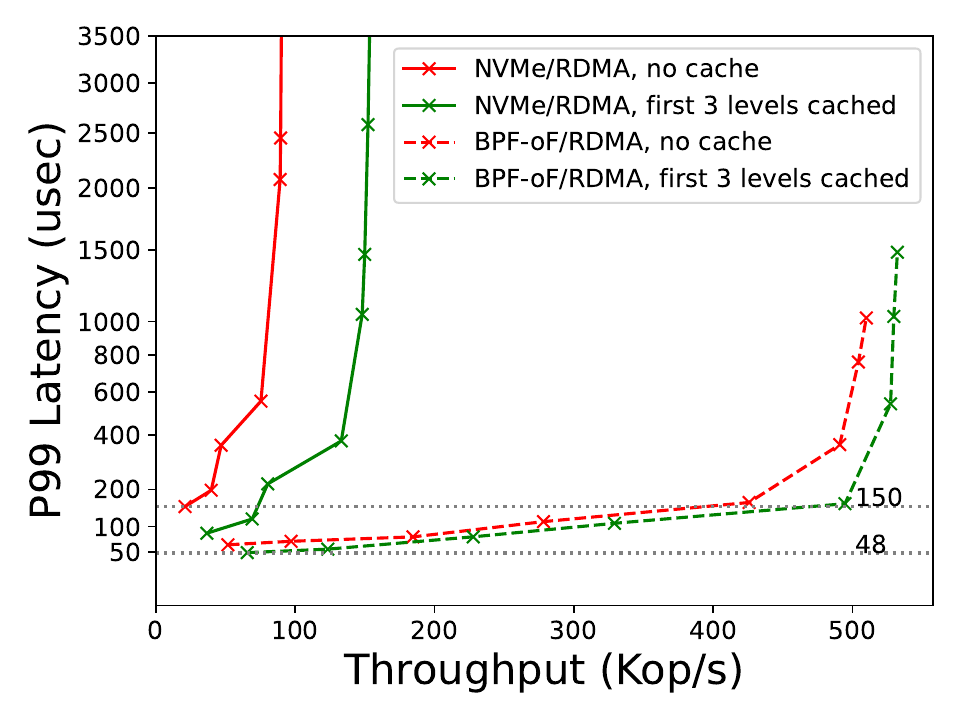}
        \vspace{-2em}
        \caption{3 cores with RDMA, Optane SSD.}
        \label{fig:3core-RDMA-Optane}
    \end{subfigure}
    \vspace{-0.5em}
    \caption{Tail latency and throughput of \name vs. NVMe-oF on BPF-KV with random 512B reads.}
    \label{fig:throughput-NAND}
    \vspace{-2em}
\end{figure}

To demonstrate the performance benefit of \name, we experiment with a simple microbenchmark using a BPF-KV instance containing 8B keys with 64B values stored on the NVMe-oF target. The application on the host issues lookups of keys at uniform random; the offered load is varied by the host.
Figure~\ref{fig:bpfkv-3core-TCP-NAND} compares the results of running \name vs. regular NVMe-oF on a NAND SSD with TCP. Beyond three cores, BPF-KV bottlenecks on the NAND disk's I/O bandwidth. The results show that on TCP, \name consistently provides about 2.6$\times$ higher throughput than NVMe/TCP when the upper levels of BPF-KV are cached, and more than 8$\times$ higher throughput when BPF-KV is run without a cache. These results are consistent when we run with different numbers of cores. Note that when no cache is used, the access pattern limits the SSD to 120K IOPS (which translates to 30K requests per second at 7 resubmissions per request), because the upper level nodes become read hot-spots in the NVMe queues.
The primary reason for the significant speedup is that \name saves significant CPU cycles in processing TCP packets, which can be used instead to generate, submit, and process additional I/O requests.

The results also show that \name provides a more modest 20-25\% improvement in unloaded \tailpct ~latency. The tail latency improvement is more modest than the throughput improvement because NAND SSD latency (90\us) is significantly higher than TCP latency for our CloudLab nodes (30\us). Therefore, since \name reduces the network roundtrips, but does not reduce the number of storage I/O requests, the unloaded latency improvement is less significant. Still, as is evident from the graphs, \name keeps the tail latency under control as the load gets higher, in contrast to the NVME-oF baseline.

Figure~\ref{fig:3core-RDMA-NAND} shows the results with RDMA and NAND. As expected, \name can accelerate NVMe-oF with RDMA, although the throughput gains are generally less significant than with TCP, since RDMA's CPU processing cost is less expensive than TCP's. The throughput peaks at around 220K because it is saturating the SSD's bandwidth.

Figures~\ref{fig:3core-TCP-Optane} and ~\ref{fig:3core-RDMA-Optane} similarly compare \name to NVMe-oF using Optane SSD. %
In this case, the throughput improvement of \name is even more significant with TCP (a speedup consistently higher than 5$\times$) since the CPU overhead of TCP is even more dominant than with NAND due to Optane's much higher IOPS (5M reads per second). Similarly, the tail latency improvement is also more significant, and is reduced by up to 3.1$\times$.
Similar to our experiments with RocksDB, in the case of \name with Optane SSD, there is almost no difference between running BPF-KV with a cache and without a cache for a uniform workload. The reason for this is that since Optane SSD is so fast (the device itself has a latency of only 3\us), the cached version of BPF-KV reduces the number I/O resubmissions by 3, saving a total of about 9\us and negligible CPU overhead. This is significantly lower than the latency (and CPU overhead) of the network, even with RDMA. %

\end{document}